\documentclass[conference]{IEEEtran}

\IEEEoverridecommandlockouts

\usepackage{amsmath}
\usepackage{amssymb}
\usepackage[dvips]{graphicx}
\usepackage{epsfig}

\newtheorem{definition}{{\bf Definition}}[section]
\newtheorem{theorem}{{\bf Theorem}}[section]
\newtheorem{lemma}[theorem]{{\bf Lemma}}
\newtheorem{corollary}[theorem]{{\bf Corollary}}

\begin{document}
\title{Analytic Controllability of Time-Dependent \\ Quantum
Control Systems}

\author{\authorblockN{Chunhua Lan}
\authorblockA{Department of Electrical \\ and Systems Engineering\\
Washington University\\
St. Louis, MO 63130-4899 \\
Email: lancb@bc.edu} \and
\authorblockN{Tzyh-Jong Tarn }
\authorblockA{Department of Electrical \\ and Systems Engineering\\
Washington University\\
St. Louis, MO 63130-4899 \\
Email: tarn@wuauto.wustl.edu}
\and
\authorblockN{Quo-Shin Chi }
\authorblockA{Department of Mathematics\\
Washington University\\
St. Louis, MO 63130-4899 \\
Email: chi@math.wustl.edu}
\and
\authorblockN{John W. Clark}
\authorblockA{Department of Physics\\
Washington University\\
St. Louis, MO 63130-4899\\
Email:jwc@wuphys.wustl.edu }}

\maketitle

\begin{abstract}
The question of controllability is investigated for a quantum
control system in which the Hamiltonian operator components carry
explicit time dependence which is not under the control of an
external agent.  We consider the general situation in which the
state moves in an infinite-dimensional Hilbert space, a drift term
is present, and the operators driving the state evolution may be
unbounded.  However, considerations are restricted by the
assumption that there exists an analytic domain, dense in the
state space, on which solutions of the controlled Schr\"odinger
equation may be expressed globally in exponential form.  The issue
of controllability then naturally focuses on the ability to steer
the quantum state on a finite-dimensional submanifold of the unit
sphere in Hilbert space -- and thus on analytic controllability. A
relatively straightforward strategy allows the extension of
Lie-algebraic conditions for strong analytic controllability
derived earlier for the simpler, time-independent system in which
the drift Hamiltonian and the interaction Hamiltonia have no
intrinsic time dependence.  Enlarging the state space by one
dimension corresponding to the time variable, we construct an
augmented control system that can be treated as time-independent.
Methods developed by Kunita can then be implemented to establish
controllability conditions for the one-dimension-reduced system
defined by the original time-dependent Schr\"odinger control
problem.  The applicability of the resulting theorem is
illustrated with selected examples.
\end{abstract}

\IEEEpeerreviewmaketitle

\section{Introduction}
Over the last two decades, quantum control has played an
important part in theoretical and experimental progress
toward the realization of laser control of chemical
reactions and the development of quantum computers
\cite {Kuriksha, Bradley, Butkovskiy_1, Butkovskiy_2, Huang_1,
Blaquiere_1, Blaquiere_2, Butkovskiy_3, Ezawa, Gordon, Rabitz,
Lloyd_1, Lloyd_2}.  Essential to this contribution has been
the integration of concepts and mathematical results from control
engineering with the fundamental principles of quantum theory.

Geometric control, a treatment of differential equations rooted in
differential geometry, unitary groups, and Lie algebras, provides
a natural mathematical basis for quantum control theory.
Explicitly or implicitly, its elements \cite{Brockett} pervade the
manipulation of quantum states in both traditional and novel
technologies.  Indeed, the field of nuclear magnetic resonance
(NMR) is largely concerned with geometric control of collections
of interacting nuclear spins \cite {Lloyd_1, Slichter, Ernst_1,
Elmsley}.  Geometric control is also a key ingredient in the
theory of quantum computation, figuring prominently in the works
of Lloyd \cite{Lloyd_3}, Deutsch \cite{Deutsch}, and Akulin
\cite{Akulin}.

In particular, Lloyd \cite {Lloyd_3} was among the first to
establish that almost all quantum logic gates are universal. More
precisely, if one has available a gate that can operate on two
qubits, plus a single-qubit operation, then an arbitrary unitary
transformation on the variables of the system can be performed
with arbitrary precision by implementing a finite sequence of
local operations.  Clark \cite{Clark} and Ramakrishna and Rabitz
\cite{Ramakrishna_1, Ramakrishna} called attention to the close
relationship between open-loop geometric quantum control methods
and the application of quantum logic gates \cite{Deutsch,
Lloyd_3}.

Following Ref.~\cite{Ramakrishna}, let us consider
differential system
\begin{equation}
{dX(t) \over dt}  =
AX(t) + \sum _{i=1}^m B_i X(t) u_i(t)\,, \quad \quad X(0)=I\,, \label {equ11}
\end{equation}
which arises both in quantum computing and molecular control.
Here, $X$ is a $N \times N$ unitary matrix ($I$ being the
corresponding identity matrix), the matrices $A$ and $B_i$, $i=1,
\ldots , m$ are $N \times N$ skew-Hermitian, and the functions
$u_i(t)$ are controls.  This equation is the law of motion of the
evolution operators which govern time development of the
$N$-dimensional vector representing a pure state of the system in
its $N$-dimensional Hilbert space.  A necessary and sufficient
condition for (\ref{equ11}) to be controllable is that the set of
all matrices generated by $A, B_i, i=1,\ldots,m$, and their
commutators (i.e., the Lie algebra generated by $A $ and $B_i$)
equals the set of all $N \times N$ skew-Hermitian matrices.
Additionally, when this condition is met, any $X$ can be attained
through some choice among the controls $u_i(t)$ restricted to
piecewise constant functions of time. In fact, the formulation
adopted by Lloyd \cite {Lloyd_3} in his universality proof
corresponds to the special case $A=0$ and $m=2$ of system
(\ref{equ11}).  Already in the 1970s, Sussmann and Jurdjevic
\cite{ Sussmann, Jurdjevic} applied Lie-group theory to obtain
rigorous results on controllability for finite-dimensional control
problems corresponding to (\ref{equ11}).

Quantum computation has mostly concerned itself with
the manipulation of discrete systems with finite-dimensional
state spaces.  However, the fundamental quantum observables
representing position and momentum, and functions thereof, are
continuous in nature. In view of recent developments in quantum
error correction \cite {Lloyd_4, Braunstein_1, Braunstein_2} and
quantum teleportation \cite {Braunstein_3, Furusawa} of
continuous variables, the potential of quantum computation
over continuous variables warrants serious investigation,
thus reopening issues of controllability on infinite-dimensional
Hilbert spaces.  Continuous quantum computers may in fact
be able to perform some tasks more efficiently than their
discrete counterparts.

As early as 1983, Huang, Tarn, and Clark (HTC) \cite {Huang_1,Huang_2}
proved a basic theorem on strong analytic controllability of quantum
systems.  This theorem explicitly embraces the case of quantum
systems whose observables are continuous quantum variables
acting on an infinite dimensional state space, but the essential
finite-dimensional results may be extracted as special cases.
Because of the difficulties caused by infinite-dimensionality and
the unboundedness of operators, an analytic domain in the
sense of Nelson \cite {Nelson} was introduced to deal with
domain problems \cite {Huang_1, Huang_2} and maintain key
features of the application of Lie algebraic methods to
finite-dimensional problems.

Infinite-dimensional control systems have been widely if not
systematically studied outside the quantum context.  Brockett
\cite {Brockett} addressed the problem of realization of
infinite-dimensional bilinear systems. Sakawa \cite{Sakawa}
introduced a method for design of finite-dimensional ${\cal H}_\infty
$ controllers for diffusion systems with bounded input and output
operators by using residual model filters.  Keulen \cite{Keulen}
designed infinite-dimensional ${\cal H}_\infty $ controllers for
infinite-dimensional systems with bounded input and output
operators by using the solutions to two kinds of Riccati equations
in an infinite-dimensional space.  Based on gap topology, Morris
\cite{Morris} constructed finite-dimensional ${\cal H}_\infty $
controllers for infinite-dimensional systems with bounded input
and output operators. Morris \cite {Morris_2} also showed that
approximations of Galerkin type can be used to design controllers
for an infinite-dimensional system.  Costa and Kubrusly \cite {Costa}
derived necessary and sufficient conditions for existence
of a state feedback controller that stabilizes a discrete-time
infinite-dimensional stochastic bilinear system and ensures that
the influence of the additive disturbance on the output is smaller
than some prescribed bound.  In Ref.~\cite{Weiss_2}, optimizability
and estimatability for infinite-dimensional linear systems are
investigated; also, a theorem on the equivalence of input-output
stability and exponential stability of well-posed infinite-dimensional
linear systems is established.  In Ref.~\cite{Wen}, the
Hilbert-space generalization of the circle criterion is used for
finite-dimensional controller design of unstable infinite-dimensional
systems.  There is also literature on absolute stability problems
and open-loop stability problems in infinite-dimensional
systems \cite{Corduneanu, Leonov, Logemann, Wexler, Rebarber}.
In addition, the spectral factorization problem plays a central
role in designing feedback control for the linear quadratic optimal
control problem in infinite-dimensional state-space systems
\cite{Callier_1, Callier_2, Staffans, Weiss}.  In contrast to
this body of work, very little has been published on controllability
for time-dependent infinite-dimensional quantum control systems.

In the microscopic world ruled by quantum mechanics, most
interesting phenomena involve change, and all real-world quantum
systems are influenced to a greater or lesser extent by
interactions with their environments.  The environment changes
with time, so the Hamiltonians used to describe these open quantum
systems are explicitly time-dependent, as in Ref.~\cite{Thorwart,
Viola}. Tailored time-dependent perturbations are used to improve
system performance \cite{Viola} in high-resolution NMR
spectroscopy, where versatile decoupling techniques are available
to manipulate the overall spin Hamiltonian \cite {Ernst_1}.
Colegrave and Abdalla studied quantum systems with a
time-dependent mass to investigate the field intensities in a
Fabry-Perot cavity \cite{Colegrave_1}. They suggested possible
applications to solid-state physics and quantum field theory \cite
{Colegrave_2}. Remaud and Hernandez \cite{Remaud} found that a
time-dependent mass parameter offers a means of simulating input
or removal of energy from the system. Implementation of controls
on these time-dependent quantum systems requires guidance from
mathematical studies of controllability for time-dependent
Hamiltonian operators.  Although the HTC theorem deals with
controllability in infinite-dimensional Hilbert space, it is
restricted to time-independent operators.  This paper explores a
more general case.  We seek an extension of the HTC theorem that
is applicable both to time-independent and time-dependent quantum
systems, as well as to systems with discrete or continuous
operators acting on finite- or infinite-dimensional state spaces.

Since this paper is aimed at an interdisciplinary readership
that includes pure quantum theorists as well as control
engineers, it is well to draw a clear distinction between time
dependence of the system arising solely from influences that
are directly under the control of an external, purposeful
agent, and time dependence that is intrinsic to the physical
system either in isolation or as embedded in a natural environment.
In the accepted terminology of control theory, which we adopt,
the former case defines a time-independent control system,
and the latter, a time-dependent system.  The issue of
controllability has received considerable attention in the
time-independent situation so identified (e.g., in
Refs.~\cite{Huang_1,Butkovskiy_3,Ramakrishna_1,Lloyd_1});
whereas relevant results for the time-dependent case are very
limited.

The time-dependent quantum control problem that we shall address is
stated formally in Sec.~2.  To cope with the unboundedness of
operators involved in the Schr\"odinger equation, an analytic
domain is introduced in Sec.~3, such that solutions of the
Schr\"odinger equation can be expressed globally in exponential
form on this domain.  In Sec.~4, we define an augmented system
in a space enlarged by one dimension, enabling its description
within the framework of time-independent control systems.
Following the pattern of Kunita's proof \cite{Kunita_1} of strong
controllability of a time-independent system, we then establish
conditions for controllability of this kind for the
one-dimension-reduced system defined by the original
time-dependent Schr\"odinger equation.  Three illustrative
applications of the theorem are presented in Sec.~5, and our
findings are reviewed in Sec.~6.

\section{Problem Formulation}

The following quantum control system is derived by
applying the geometric quantization method \cite{Santilli}
to a classical bilinear control system \cite{Tarn,Huang_2}:
\begin{equation}
\begin{split}
 i\hbar \frac{\partial }{\partial t} \psi (t) &= \left[ H_0^\prime
(t) + \sum_l u_l (t) H_l^\prime (t)\right] \psi (t), \\
& \psi (t_0)=\psi_0 \,.
\end{split}
\end{equation}
Here, $H_0^\prime (t)$, and the $ H_l^\prime (t)$ with $l=1,2,
\ldots , r$, are Hermitian operators on a unit sphere $S_{\cal H}$
of Hilbert space, the $u_l(t), \, l=1, \ldots , r$ are restricted
to piecewise-constant real functions of time, and $\psi (t)$
denotes a quantum state belonging to $S_{\cal H}$.  In physical
language, $H_0^\prime $ is the unperturbed or autonomous
Hamiltonian, and the $H_l^\prime $ are interaction Hamiltonians.
It is the coefficients $u_l(t)$ that are subject to purposeful
control by an agent external to the system, within the specified
class of functions.
Setting $\hbar =1$ and dividing $H_0^\prime (t)$ and the $H_l^\prime (t)$
by $i$, we arrive at a more familiar control form,
\begin{equation}
\begin{split}
\frac{\partial }{\partial t} \psi (t) & = \left[H_0 (t) + \sum_l u_l (t)
H_l (t)\right] \psi (t)\,, \\
& \psi (t_0) = \psi_0 \in S_{\cal H} \,, \label{equ411}
\end{split}
\end{equation}
where the $H_i (t)$, $i=0,1,2, \ldots , r$, are skew-Hermitian
operators on $S_{\cal H}$.  From the standpoint of systems
engineering, $H_0(t)$ is called the drift term in
Eq.~(\ref{equ411}) because no control function directly modifies
its action. Importantly, we depart from previous studies of
quantum controllability in allowing the Hamiltonian operators
$H_i(t)$ to their own carry explicit time dependence, which is
assumed to be inherent in the physical structure of the system and
therefore beyond the control of any external agent. The operators
$H_i(t)$ are the counterparts of the structural matrices involved
in standard formulations of linear control theory.

For the system (\ref{equ411}), we know from arguments presented in
Ref.~\cite{Huang_1} that the transitivity of states on $S_{\cal H}$
requires an infinite sequence of control manipulations within
the control set $\{u_l(t)\}$ of piecewise-constant real
functions.  Clearly, such a process is strictly meaningless
in practice, although under certain conditions it may be
possible to find a finite series of control operations that
approach the desired target state arbitrarily closely.
Even so, we are naturally directed to consider the issue
of controllability on a {\it finite-dimensional} submanifold
of the unit sphere $S_{\cal H}$, for which in turn a finite-dimensional
tangent space is generated by $H_0 (t) \psi (t), \ldots , H_r (t) \psi (t)$.

Accordingly, our attention focuses on a finite-dimensional
submanifold $M \subset S_{\cal H}$, on which the following
dynamics prevail
\begin{align}
& \frac{\partial }{\partial t} \psi (t) = \left[H_0 (t)
+ \sum_l u_l(t) H_l (t)\right]\psi (t)\,, \nonumber \\
& \psi (t_0) = \psi _0, \, \psi (t) \in M,\, \forall t \geq
t_0 \,,\label{equ421}
\end{align}
Thus, instead of studying controllability on $S_{\cal H}$, we
consider controllability on $M \subset S_{\cal H}$.  On the
submanifold $M$, the inherited topology of $S_{\cal H}$
still applies; hence it is paracompact and connected.

For system (\ref{equ421}), we have available a set of vector fields
$O(M)$ composed of skew-Hermitian operators on $M$ with Lie
algebra defined by $O(M) = {\cal L} \{H_0, \ldots , H_r \}$.
Let $V$ be a subset of $O(M)$.  The Lie algebra generated by $V$
is denoted by ${\cal L}(V)$.  The restriction of ${\cal L}(V)$
to a point $\psi$ on $M$, which is a tangent subspace of $TM_\psi$
at $\psi$, is written as
\begin{equation}
{\cal L} (V)(\psi ) = \{ Y\psi  | Y \in {\cal L}(V) \} \subset
TM_\psi\,,
\end{equation}
while
\begin{equation}
\tilde {\cal L}(V) = \{ {\cal L}(V)\psi  | \psi \in M\}
\end{equation}
defines an involutive differential system.  A vector field $X$ is
said to belong to $\tilde {\cal L}(V)$ if $X(\psi ) \in
\tilde{\cal L}(V)(\psi )$ holds for all $\psi \in M$.

\section{Selecting the Domain}

Recognizing that operators in quantum mechanics are in general
unbounded, we need to find a domain on which exponentiations of
the operators entering the system (\ref{equ421}) converge.
To this end, we introduce the so-called analytic domain conceived
by Nelson \cite {Nelson}, a dense domain invariant under the action
of the operators in system (\ref{equ411}). The solution of the
Schr\"odinger equation can be expressed globally in exponential form
on this domain, which is also invariant under the action of the
exponentiations of the operators $H_i$.
\begin{definition}
If $H$ is an operator on the state space $\cal H$, we call an
element $\omega$ of $\cal H$ an analytic vector for $H$ in case the
series expansion of $\exp(Ht) \omega$ has a positive radius of absolute
convergence, that is, provided
\begin{equation}
\sum^\infty_{n=0}  \frac{||H^n \omega ||}{n!} s^n < \infty
\end{equation}
for some $s>0$.
\end{definition}
If $H$ is a bounded operator, then every vector in $\cal H$ is
trivially an analytic vector for $H$.

The corresponding definition of analytic vectors for a Lie
algebra of operators runs as follows \cite{Nelson, Barut}:

\begin{definition}
A vector $\omega \in \cal H$ is said to be an analytic vector for the
whole Lie algebra $\cal L$ if for some $s > 0$ and some linear
basis $\{ H_1, \ldots , H_d \} $ of the Lie algebra, the series
\begin{equation}
\sum_{n=0}^\infty \frac{1}{n!}  \sum _{1 \leq i_1, \ldots , i_n
\leq d} ||H_{i_1} \ldots H_{i_n} \omega || s^n
\end{equation}
is absolutely convergent.
\end{definition}

The concept of analytic vectors is especially useful for our
purposes, since for certain types of unbounded operators
they form a dense set in the Hilbert space.  In fact, the
set of all analytic vectors for a Lie algebra $\cal L$ forms
an analytic domain in the following sense \cite{Nelson,Barut}.

\begin{definition}
Let $\cal L$ be the Lie algebra generated by the skew-Hermitian
operators ${H_0, \ldots , H_r}$ on a unit sphere $S_{\cal H}$
of Hilbert space.   An analytic domain ${\cal D}_A$ is said to
exist for the $H_i$, $i=0,1,\ldots r$, if (i) there exists a common
dense invariant subspace ${\cal D}_A \subset \cal H$ on which the
corresponding unitary Lie group $G$ can be expressed locally
in exponential form with Lie algebra $\cal L$, (ii) ${\cal D}_A $
is invariant under $G$ and $\cal L$, and (iii) on ${\cal D}_A $,
elements of $G$ can be extended globally to all $t \in {\mathbb R}^+$.
\label{def421}
\end{definition}

We now state Nelson's fundamental theorem, which provides conditions
under which a Lie algebra $\cal L$ defined by a set of skew-Hermitian
operators can be associated with a unitary group $G$ having
$\cal L$ as its Lie algebra.

\begin{theorem} (Nelson)
Let $\cal L$ be a Lie algebra of skew-Hermitian operators in a
Hilbert space $\cal H$ which have a common invariant dense domain
${\cal D}_A$.  Let $X_1, \ldots , X_d$ be an operator basis for $\cal
L$. If $T = X_1^2 + \ldots + X_d^2$ is essentially self-adjoint,
then there is  a unique unitary  group $G$ in $\cal H$ with Lie
algebra $\cal L$.  Let $\overline T$ denote the unique
self-adjoint extension of $T$. Then the analytic vectors of
$\overline T$ are analytic vectors for the whole Lie algebra
${\cal L}$ and form a set invariant under $G$ and dense in ${\cal
H}$. \label{theo342}
\end{theorem}

Accordingly, on the analytic domain ${\cal D}_A$, the Lie algebra and
its unitary Lie group are related through the familiar exponential
formula. The Lie algebra is composed of skew-Hermitian
operators which are vector fields defined on ${\cal D}_A \cap
S_{\cal H}$.  By property (iii) of the definition \ref{def421} of
the analytic domain, these vector fields on ${\cal D}_A \cap S_{\cal
H}$ are complete.  Moreover, owing to the skew-Hermiticity of
the operators $H_i$ of system (\ref{equ411}), the corresponding
transformation groups, taking a given state on $S_{\cal H}$ to another
state on $S_{\cal H}$, are unitary.  This feature guarantees
preservation of the norm of quantum states, as required for the
statistical interpretation of quantum mechanics.

In fact, Nelson's theorem only provides sufficient conditions for
the important properties it yields.  With this in mind, we shall
assume an analytic domain ${\cal D}_A$ exists {\it without} explicitly
imposing the conditions stated in this theorem, a stance also
adopted in Ref.~\cite{Huang_1}  This strategy clearly
implies that the existence of such a domain must be established
explicitly prior to application of the controllability results
to be derived in the following sections.

We are now prepared to adapt the concept of controllability to
problems involving unbounded operators.
\begin{definition}
For system (\ref{equ411}), if ${\cal D}_A$ exists for $\cal L$,
and if for any $\psi_0$ and $\psi_f \in {\cal D}_A \cap S_{\cal
H}$ there exist control functions $u_1(t), \ldots , u_r(t)$, and a
time $t_f$ [resp.\ $\forall t_f$] such that the solution of
control system (\ref{equ411}) satisfies $\psi (t_0) = \psi_0$,
$\psi (t_f) = \psi_f$, and $ \psi(t) \in {\cal D}_A \cap S_{\cal
H}$, where $t_0 \leq t \leq t_f$, then the system is called
analytically controllable [resp.\ strongly analytically
controllable] on $S_{\cal H}$; moreover we then say that the
corresponding unitary Lie group is analytically transitive on
$S_{\cal H}$.
\end{definition}

As has been argued, the more pertinent concept is controllability
on the submanifold $M$ of $S_{\cal H}$.  By assumption, $M \cap
{\cal D}_A $ is dense in $M$, while $\dim (M \cap {\cal D}_A ) =
\dim M =m $.  Denoting the tangent space of $M \cap {\cal D}_A$ at
$\psi $ by $TM_\psi = {\mathcal L} \{ H_0  , \ldots , H_r  \}
\psi$, the tangent bundle of the system (\ref{equ421}) is given by
$T(M\cap {\cal D}_A) = \cup_{\psi \in {M \cap {\cal D}_A }}
TM_\psi $.

Let $R_t(\psi )$ denote the set of all points that are reachable
from $\psi $ at time $t$.  The set $R (\psi ) = \bigcup _{t>t_0}
R_t(\psi )$ is then reachable from $\psi $ at some time
greater than $t_0$.  We say that system (\ref{equ421}) is analytically
controllable on $M$ if $R (\psi ) = M \cap {\cal D}_A , \,
\forall \psi   \in M \cap {\cal D}_A $, and that the system is strongly
analytically controllable on $M$ if $R_t (\psi ) = M \cap
{\cal D}_A, \forall t > t_0, \, \forall \psi  \in M \cap {\cal D}_A$.

\section{Controllability of Time-dependent Quantum Control Systems}

\subsection{Reformulation as a Time-independent Augmented System}

Most of the methods developed for determining controllability
of time-independent bilinear or nonlinear systems
\cite{Kunita_2, Isidori, Huang_1, Huang_2, Clark_1, Clark_2}
cannot be applied directly to the time-dependent bilinear control
problem studied here, since these approaches rely upon the
following property.  Let $Y_t(\varphi)$ be an integral curve of
the time-independent tangent vector $Y$ starting from point
$\varphi$ and $t \in [t_0, t_0+t_f]$, and let $cY_t(\varphi)$ be an
integral curve of the tangent vector $cY$ starting
from $\varphi$ and $t \in [t_0, t_0+t_f/||c||]$; then the integral
curves $Y_t(\varphi)$ and $cY_t(\varphi)$ coincide.  This property
holds for all time-independent tangent vectors, but it generally
fails for time-dependent tangent vectors.

However, recognizing that this feature has been instrumental
to controllability proofs for nonlinear systems, we recast
the system (\ref{equ421}) as a time-independent problem so that
it can once again be exploited.  Reformulation of the original
problem is accomplished by regarding the time variable $t$ as
an additional parameter in the specification of the system state,
supplementing the state vector $\psi$.  Thus the state of the
extended system is expressed as
\begin{equation}
\xi = \left ( \begin{array}{c}
               t +t_0 \\
               \psi
               \end{array}  \right ) \,.
\label{equdef441}
\end{equation}
Making the corresponding extension of the manifold $M$, we
form an augmented $(m+1)$-dimensional manifold defined by
\begin{equation}
 N = \left \{ \begin{array}{c}
               {\mathbb R} \\
               M \cap {\cal D}_A
               \end{array}  \right \},
\label{defN}
\end{equation}
where ${\mathbb R}$ is the real line.  Next we define augmented
vector fields $W_l$ by
\begin{equation}
\begin{split}
& W_0 (\xi ) = \left [ \begin{array}{c}
               1 \\
               H_0(t+t_0) \psi (t+t_0)
               \end{array}  \right ], \\
& W_l (\xi ) = \left [ \begin{array}{c}
               0 \\
               H_l(t+t_0) \psi (t+t_0)
               \end{array}  \right ],
\label{equdef442}
\end{split}
\end{equation}
with $l= 1,2, \ldots , r$.  Obviously, the $W_l$, with $l=0, 1,
\ldots , r$, depend on both $t$ and $\psi$, i.e., the $W_l$ now
depend on the state $\xi$ defined by Eq.~(\ref{equdef441}).

The time-dependent control system (\ref{equ421}) has thereby been
reformulated as an augmented system of time-independent form.
Explicitly,
\begin{align}
& \frac{\partial \xi (t) }{\partial t} = \left[ W_0 (\xi )
+ \sum_l u_l(t) W_l(\xi)\right] \,, \label{equ431} \\
& \xi(0) = \eta = \left ( \begin{array}{c}
                      t_0 \\
                      \psi (t_0)
                      \end{array}  \right ) = \left ( \begin{array}{c}
                                                      t_0 \\
                                                      \psi_0
                                                      \end{array} \right )\,,
                      \nonumber \\
& \forall t \geq 0,\, \psi_0 \in M \cap {\cal D}_A,\, \xi \in N \,,
\nonumber
\end{align}
where $N$ is the $n=(m+1)$-dimensional manifold constructed in
Eq.~(\ref{defN}) and $M$ is now viewed as a one-dimension-reduced
manifold of the augmented system.  As always, the controls $u_l(t)$,
with $l= 1, \ldots , r$, are piecewise-constant real functions of
time $t$.

It is convenient to employ $t+t_0$ instead of $t$ in definitions
(\ref{equdef441}) and (\ref{equdef442}), thereby setting the starting
time at zero for the augmented system (\ref{equ431}).  Since the
latter system is time-independent by construction, this can be
done without affecting its trajectory.  Thus, if the time for the
augmented system is $t$, then the time for the original system
(\ref{equ421}) is $t+t_0$.  Standard differential equation techniques
can evidently be employed to analyze the behavior of the augmented
system on the manifold $N$, and the results will reflect the behavior
of the original system on manifold $M$.

We note peripherally that system (\ref{equ431}) is in a decomposed
form in the sense of Ref.~\cite{Isidori}, where several
theorems were developed for decomposition of nonlinear control
systems.  However, these theorems do not specify reachable sets,
so they cannot be applied here to obtain controllability results.

Reachable sets ${\hat R}_t(\eta)$ and ${\hat R}(\eta)$ are defined
for the augmented system (\ref{equ431}) in just the same manner as
for system (\ref{equ421}).  From the work of Huang, Tarn, and
Clark \cite{Huang_1} based on the results of Chow \cite{Chow},
Sussmann and Jurdjevic \cite{Sussmann}, and Kunita
\cite{Kunita_1,Kunita_2}, it is to be expected that the issue of
analytic controllability will hinge on the relationships among
certain Lie algebras generated by the vector fields involved in
the control system (\ref{equ421}) or its augmented counterpart
(\ref{equ431}). For the latter problem, these Lie algebras are
specified by ${\hat {\cal A}} = { {\cal L}} \{ W_0, \ldots , W_r
\}$, $ \hat{{\cal B}} = { {\cal L}} \{W_1, \ldots , W_r \}$, and
${\hat {\cal C}} = {\cal L} \{ {\rm ad}_{W_0}^m W_l,\, l=1, \ldots
,r,\, m=0, \ldots , \infty \}$.  By definition, ${\rm ad}_{W_0}^m
W_l$ is built from repeated commutators of $W_0$, present in
${\hat {\cal A}}$ but not ${\hat {\cal B}}$, with any and all of
the $W_l$ present in ${\hat {\cal A}}$ or ${\hat {\cal B}}$;
clearly,
\begin{equation}
{\hat {\cal B}} \subset {\hat {\cal C}}
\subset {\hat {\cal A}} \,.
\label{equnew1}
\end{equation}
For future reference we note (in particular) that
the restriction of ${\hat {\cal B}}$ to a point $\psi $ on $N$,
which is a tangent subspace of $TN_{\psi }$ at $\psi $, is written as
\begin{equation}
{\hat {\cal B}}(\psi )=\{Y(\psi )| Y \in  {\hat {\cal B}}\} \subset TN_{\psi },
\end{equation}
and in turn that
\begin{equation}
\widetilde{\hat {\cal B}} = \{ {\hat {\cal B}}(\psi ) | \psi \in N
\}
\end{equation}
is an involutive differential system.

\subsection{Controllability of the Augmented System}

We must still face the situation that standard controllability results
\cite {Kunita_2, Isidori, Huang_1, Huang_2, Clark_1, Clark_2},
derived for time-independent systems, cannot be carried over directly
to our problem as reformulated in the preceding subsection, since
derivation of these results employs the vector-space property of
the tangent space.  Specifically, it is required that if $Y$ is an
acceptable tangent vector, then so is $cY$, where $c$ is an
arbitrary constant.  But in our case, once the first component of
a tangent vector of the augmented manifold is fixed at unity,
it is not possible for both $Y$ and $cY$, with $c \neq 1$, to be
available tangent vectors.  However, with the aid of
a result of Kunita \cite {Kunita_1}, we may nevertheless establish
one-dimension-reduced controllability of the augmented system;
that is, we may prove strong analytic controllability of the original
system since it is not necessary to control the time dimension.

First, let us identify certain properties of the reachable set
${\hat R}_t(\eta)$ that will be useful in proving strong analytic
controllability.

\begin{theorem} \cite{Sussmann, Kunita_1}
Assume that the Lie algebra $\hat{\cal C}$ is locally finitely generated,
and let $I(\eta )$ be the maximal connected integral manifold of
$\hat{\cal C}$ containing the point $\eta $.  Then ${\hat R}_t( \eta ) \subset
\alpha_t^0 (I(\eta ))$, where $\alpha_t^0 $ is the integral curve whose
vector field is $W_0$.  Furthermore, the interior of ${\hat R}_t(\eta )$
with respect to the topology of $\alpha_t^0 (I(\eta ))$ is dense
in ${\hat R}_t(\eta )$. \label{the42}
\end{theorem}

A key relationship between the interior of the reachable set
${\hat R}_t(\eta )$ of the augmented system at time $t$ and the interior
of its closure is provided by the following lemma.

\begin{lemma}
\begin{equation}
{\rm int}({\rm cl}\,{{\hat R}_t(\eta )}) = {\rm int}\, {\hat
R}_t(\eta )\,. \label{lemthe43}
\end{equation}
\end{lemma}
\noindent {\bf Proof: } Let $\chi \in {\rm int} ({\rm cl}\,
{{\hat R}_t(\eta)})$ and let $S_\epsilon (\chi)$ be the set of
all $\chi^\prime $ such that $\chi$ is reachable from $\chi^\prime $
within time $\epsilon > 0$. Then $S_\epsilon (\chi)$ is the reachable
set within time $\epsilon > 0$ for the dual control system
\begin{equation}
\frac{\partial \upsilon }{\partial t} = -\left[ W_0(\upsilon)
                          + \sum_l u_l(t) W_l(\upsilon)\right]\,.
\end{equation}
Theorem 4.1 implies that ${\rm int}\, S_\epsilon (\chi)$ is dense
in ${\rm cl}\, {S_\epsilon (\chi)}$, and ${\rm int}\,{\hat
R}_t(\eta )$ is dense in ${\rm cl}\, {{\hat R}_t(\eta )}$.  Since
$\chi \in {\rm cl}\, {S_\epsilon (\chi)}$, we know that
\begin{equation}
{\rm cl}\, {S_\epsilon (\chi)} \cap {\rm int}({\rm cl}\,
{{\hat R}_t(\eta )})
\not= \emptyset
\end{equation}
and hence that
\begin{equation}
{\rm int}\, S_\epsilon (\chi) \cap {\rm int} ({\rm cl}\,
{{\hat R}_t(\eta )}) \cap {\hat R}_t(\eta ) \not=  \emptyset \,.
\end{equation}
If $\zeta$ belongs to the latter intersection, then $\zeta$ is
reachable from $\eta $ using time $t$, and $\chi$ is reachable from
$\zeta$ in elapsed time less than or equal to $\epsilon $.  Therefore,
$\chi$ is reachable from $\eta $ in elapsed time between $t$
and $ t + \epsilon$.  This argument holds for any $t >0 $
and any $\epsilon >0$.  Letting $\epsilon  \rightarrow  0$, we
conclude that $\chi$ is reachable from $\eta $ in time $t$,
so $\chi \in {\hat R}_t(\eta )$.  Thus,
$$
{\rm int}({\rm cl}\, {{\hat R}_t(\eta )})
\subset {\hat R}_t(\eta )
\Longrightarrow {\rm int}( {\rm cl}\, {{\hat R}_t(\eta )})
\subset {\rm int}\, {\hat R}_t(\eta )\,.
$$
But clearly ${\rm int}\, {\hat R}_t(\eta ) \subset {\rm int} ({\rm
cl}\, {{\hat R}_t(\eta )})$ and the statement (\ref{lemthe43})
follows.

From the control-theoretic perspective, the drift term is
undesirable because no control is present to influence or
remove its effect.  It is therefore of strategic value to consider
a suitably modified control system, called the auxiliary system,
that will serve as a bridge to an effective controllability
analysis of the augmented system.  Let $e_0, e_1, \ldots , e_r$
be unit vectors in ${\mathbb R}^{r+1}$; in particular, let $e_i =
(0, \ldots , 0, 1, 0, \ldots , 0)$, in which only the $(i+1)^{th}$
element is unity and the others are zero.  Denote by ${\cal U}_0$
the set of controls $u(t) = \left(u_0(t), \ldots , u_r(t)\right)$
composed of piecewise-constant functions $u_i(t)$ taking the
values $e_0, \pm e_1, \ldots , \pm e_r$ only.  Consider then the
control system expressed in the form
\begin{equation}
\frac{\partial \xi }{\partial t} =  u_0(t) W_0(\xi) + \sum_l u_l(t) W_l(\xi)
\,, \quad \xi (t_0) = \eta \,,\label{equ432}
\end{equation}
where $u(t) \in {\cal U}_0$.  The solution of this system may
be written as
\begin{equation}
\alpha_t = \alpha_{t_k}^{i_k} \cdots \alpha_{t_j}^{i_j} \cdots
\alpha_{t_1}^{i_1},  \label{equexpre1}
\end{equation}
where $k$ is a positive integer and where $\alpha_{t_j}^{i_j}$ is
the integral curve of $W_{i_j}$ with $i_j = 0, 1, \ldots , r$,
$j=1,\ldots,k$, and $k$ a positive integer. The times $t_j$
satisfy $t_j \geq 0$ if $i_j =0$, $t_j \in {\mathbb R}$. We denote
by ${\hat R}_t^0(\eta)$ the reachable set of the auxiliary system
corresponding to the total time $t$ since time zero, over which
the control function $u_0(\cdot)$ is nonzero; the reachable set of
the auxiliary system is then ${\hat R}^0(\eta )= \bigcup_{t > 0}
{\hat R}_t^0(\eta)$. Theorem \ref{the42} is valid for this control
system \cite {Sussmann}.

The following notations are convenient:
\begin{align*}
{\rm Exp}\, {\hat {\cal L}} = & \mbox{ the group of
diffeomorphisms generated}  \mbox{ by } \\
& \mbox{ the }\alpha_t^i,\, t \in {\mathbb R},\, i = 0, \ldots ,
r,  \mbox{ where } \alpha_t^i
\mbox{ is } \mbox{ an }\\
& \mbox{ integral curve of } W_i \,,  \\
 ({\rm Exp}\, {\hat {\cal L}})_+  = & \mbox{ the semigroup of
diffeomorphisms }  \mbox { generated }\\
& \mbox{ by } \alpha_t^0,\, t \geq 0, \mbox{ and the }
\alpha_t^l,\, \mbox{ with } t \in {\mathbb R} \\
& \mbox{ and }
l= 1, \ldots , r\,, \\
({\rm Exp}\, {\hat{\cal L}})_t  = & \mbox{ the subset of } ({\rm
Exp}\,{\hat{\cal L}})_+ \mbox{ generated }  \mbox { by }\\
& \alpha_{t_k}^{i_k} \cdot
 \ldots \cdot \alpha_{t_1}^{i_1} \,, \mbox{ with } \sum _{j=1}^k t_j \cdot 1_{\{i_j =0\}} = t
 \,.
\end{align*}
To clarify the meaning of the last line, we note that when the
index $j$ is such that $i_j=0$, we have $u_0= 1$ (and all the
other $u_i=0$), so $W_0$ is ``turned on'' and does play a role as
an active vector field or tangent vector.  Conversely, for indices
$j$ such that $i_j \neq 0$, the factor $u_0$ multiplying $W_0$ in
system (\ref{equ432}) vanishes, and $W_0$ plays no role.  The sum
appearing in the definition of $({\rm Exp} \, {\hat{\cal L}})_t$
gives the total time over which $W_0$ is active in the system
dynamics.

From Chow's theorem \cite{Chow,Sussmann}, it is known that the group
${\rm Exp}\, {\hat{\cal L}}$ acts transitively on the manifold $N$ when
$\dim {\hat{\cal L}}\{W_0, W_1, \ldots , W_r \}= \dim N$, i.e.,
we know that  $\{\alpha (\eta ) | \alpha \in {\rm Exp}\,
{\hat{\cal L}} \} = N$ for any $\eta \in  N$. On the other hand, the
reachable set at time $t$ for the auxiliary system (\ref{equ432})
is ${\hat R}_t^0(\eta ) = \{ \alpha (\eta ) | \alpha \in
({\rm Exp}\,{\hat{\cal L}})_t \}$.  (It is to be noted that in the
present context $t$ is the total time over which $W_0$ has been
active since time zero, which is generally not equal to the actual
elapsed time, since $W_0$ may be turned off over certain intervals.)

\begin{lemma}
\begin{equation}
{\rm cl}\, {{\hat R}_t(\eta )} = {\rm cl}\, {{\hat R}_t^0(\eta )}
\label{equlem41} \,.
\end{equation}
\label{lem41}
\end{lemma}
We may gain intuitive understanding of this lemma by analyzing
a simple example.

\noindent
{\it Example}.  \quad Let us compare the control system
\begin{equation}
{d \over dt} {\left ( \begin{array}{c} x \\y \end{array} \right )} =
\left ( \begin{array}{c} 1 \\0 \end{array} \right )
+ u \left ( \begin{array}{c} 0 \\1 \end{array} \right ) \,,
\label{equ436}
\end{equation}
wherein $u \in {\mathbb R}$, with the system
\begin{equation}
{d \over dt} {\left ( \begin{array}{c} x \\y \end{array} \right )} =
u_0\left ( \begin{array}{c} 1 \\0 \end{array} \right ) + u_1 \left
( \begin{array}{c} 0 \\1 \end{array} \right ) \,,
\label{equ437}
\end{equation}
wherein $(u_0, u_1) \in \{(0, \pm 1), (1,0) \}$. Clearly, the
first of these corresponds to the augmented system, and the second
to the auxiliary system.  Let ${\hat R}_t(\eta )$ and ${\hat
R}_t^0(\eta )$ denote respectively the reachable sets of systems
(\ref{equ436}) and (\ref{equ437}), staring from the state $\eta$.
While stopping short of rigorous argument, explicit computation
will be used to reveal the pertinent relationship between ${\rm
cl}\, {{\hat R}_t(\eta )}$ and ${\rm cl}\,{{\hat R}_t^0(\eta )}$.

First consider the integral curve
\begin{equation}
\alpha_t(\eta ) =\left ( \begin{array}{c} 0 \\1
\end{array} \right )_{t_1} \cdot \left ( \begin{array}{c} 0 \\-1
\end{array} \right )_{t_2} \cdot \left ( \begin{array}{c} 1 \\0
\end{array} \right )_{t} \in {\hat R}_t^0(\eta )  \,,
\end{equation}
and for $n= 1, 2, 3, \ldots$ form a series of integral curves
$\beta_t^n(\eta) \in {\hat R}_t(\eta)$ defined by
\begin{equation}
\begin{split}
 \beta_t^n(\eta )  =&
  \left ( \left (
\begin{array}{c} 1\\0
\end{array} \right )+n\left ( \begin{array}{c} 0 \\1
\end{array} \right ) \right )_{\frac{t_1}{n}}  \\
& \cdot \left ( \left (
\begin{array}{c} 1\\0
\end{array} \right )+n\left ( \begin{array}{c} 0 \\-1
\end{array} \right ) \right )_{\frac{t_2}{n}}  \cdot \left ( \begin{array}{c} 1
\\0\end{array} \right )_{t-\frac{t_1}{n}-\frac{t_2}{n}} \,.
\end{split}
\end{equation}
As $n$ goes to $\infty $, we find
\begin{equation}
\beta_t^n (\eta ) \rightarrow
\left (
\begin{array}{c} 0 \\1
\end{array} \right )_{t_1} \cdot \left ( \begin{array}{c} 0 \\-1
\end{array} \right )_{t_2} \cdot \left ( \begin{array}{c} 1 \\0
\end{array} \right )_{t}\,,
\end{equation}
that is, $\beta_t^n(\eta) \rightarrow \alpha_t(\eta )$.
Hence $\alpha_t(\eta ) \in {\rm cl}\, {{\hat R}_t(\eta )}$.

On the other hand, consider
\begin{equation}
\begin{split}
\beta_t(\eta )  = & \left ( \left (
\begin{array}{c} 1\\0
\end{array} \right )+m_1\left ( \begin{array}{c} 0 \\1
\end{array} \right ) \right )_{t_1} \cdot \left ( \begin{array}{c} 1 \\0
\end{array} \right )_{t_2} \\
& \cdot \left ( \left ( \begin{array}{c} 1\\0
\end{array} \right )+m_2\left ( \begin{array}{c} 0 \\-1
\end{array} \right ) \right )_{t_3} \in {\hat R}_t(\eta )\,,
\end{split}
\end{equation}
where $m_1, m_2 \in {\mathbb R}$ and $t= t_1+t_2+t_3$, and construct
\begin{equation}
\alpha_1^n = \left [ \left ( \begin{array}{c} 1 \\ 0 \end{array}
\right )_{\frac{t_1}{n}} \cdot m_1 \left ( \begin{array}{c} 0 \\1
\end{array}\right )_{\frac{t_1}{n}} \right ]^n \,,
\end{equation}
again for $n=1,2,3, \ldots$.
Applying the Baker-Campbell-Hausdorff formula, it straightforward
to show that
\begin{equation}
\begin{split}
\lim_{n\rightarrow \infty } \alpha_1^n  & =  \lim_{n \rightarrow
\infty } \left \{ \left ( \left (
\begin{array}{c} 1\\0
\end{array} \right )+m_1\left ( \begin{array}{c} 0 \\1
\end{array} \right ) \right )_{t_1} \right. \\
& \quad \left.  + \frac{t_1^2}{2n} m_1 \left [ \left (
\begin{array}{c} 1
\\0
\end{array} \right ), \left (
\begin{array}{c} 0 \\1 \end{array} \right ) \right ]
+O(\frac{1}{n^2}) \right \} \\
& = \left ( \left (
\begin{array}{c} 1\\0
\end{array} \right )+m_1\left ( \begin{array}{c} 0 \\1
\end{array} \right ) \right )_{t_1}\,.
\end{split}
\end{equation}
Similarly, let
\begin{equation}
\alpha_3^n = \left [ \left (
\begin{array}{c} 1 \\0 \end{array} \right )_{\frac{t_3}{n}} \cdot m_2 \left (
\begin{array}{c} 0 \\-1 \end{array}\right )_{\frac{t_3}{n}} \right
]^n
\end{equation}
and employ the Baker-Campbell-Hausdorff formula to obtain
\begin{equation}
\begin{split}
\lim_{n\rightarrow \infty } \alpha_3^n  & = \lim_{n \rightarrow
\infty } \left \{ \left ( \left (
\begin{array}{c} 1\\0
\end{array} \right )+m_2\left ( \begin{array}{c} 0 \\-1
\end{array} \right ) \right )_{t_3} \right.  \\
 & \quad  \left.  + \frac{t_3^2}{2n} m_2 \left [ \left (
\begin{array}{c} 1
\\0
\end{array} \right ), \left (
\begin{array}{c} 0 \\-1 \end{array} \right ) \right ]
+O(\frac{1}{n^2}) \right \} \\
& = \left ( \left (
\begin{array}{c} 1\\0
\end{array} \right )+m_2\left ( \begin{array}{c} 0 \\-1
\end{array} \right ) \right )_{t_3}\,.
\end{split}
\end{equation}
Obviously
\begin{equation}
 \alpha_1^n \cdot \left (
\begin{array}{c} 1 \\0 \end{array} \right )_{t_2} \cdot \alpha
_3^n \in {\hat R}_t^0(\eta )\,,
\end{equation}
and we find that
\begin{equation}
\begin{split}
\lim_{n\rightarrow \infty } & \alpha_1^n  \left (
\begin{array}{c} 1 \\0 \end{array} \right )_{t_2}  \alpha
_3^n  = \left ( \left (
\begin{array}{c} 1\\0
\end{array} \right )+m_1\left ( \begin{array}{c} 0 \\1
\end{array} \right ) \right )_{t_1} \\
& \cdot \left ( \begin{array}{c} 1 \\0 \end{array} \right )_{t_2}
\cdot \left ( \left (
\begin{array}{c} 1\\0
\end{array} \right )+m_2\left ( \begin{array}{c} 0 \\-1
\end{array} \right ) \right )_{t_3}  = \beta_t(\eta )\,.
\end{split}
\end{equation}
Therefore $\beta_t(\eta ) \in {\rm cl}\, {{\hat R}_t^0(\eta )}$.

Now let us proceed with the proof of Lemma \ref{lem41}, showing
first that ${\rm cl}\, {{\hat R}_t^0(\eta )} \subseteq {\rm cl}\,
{{\hat R}_t(\eta )}$. Consider that $\alpha_t(\eta ) \in  {\hat
R}_t^0(\eta )$ is expressible in the form of $\alpha_{t_k}^{i_k}
\cdots \alpha_{t_1}^{i_1}(\eta )$, where $ t= \sum _{j=1}^k t_j
\cdot 1_{\{i_j =0\}}$.  With the guidance of the example above, a
sequence of controls $u^{(n)}(\cdot)$ associated with the
diffeomorphism of this form is constructed as follows.  For an
arbitrary positive integer $n$ such that $n t_m \geq \sum_{i_j
\neq 0}|t_j|$, where $m$ is the last subscript $j$ such that
$i_j=0$, let
\begin{equation}
t_m^{(n)} = t_m -\frac{\sum_{i_j \neq 0}|t_j|}{n}\,.
\end{equation}
Define real numbers $s_1^{(n)}, \ldots , s_k^{(n)}$, ordered so that
$ 0 \leq s_1^{(n)} \leq s_2^{(n)} \leq \ldots \leq s_k^{(n)}$,
by
\begin{equation}
\begin{array}{l}
\begin{array}{l@{\quad = \quad }l@{\quad \mbox {if }\quad }l}
s_1^{(n)} & |t_1| & i_1 = 0\,,\\
          & \frac{1}{n} |t_1| & i_1 \not= 0\,,
\end{array}
\\
\begin{array}{l@{\quad = \quad }l@{\quad \mbox {if }\quad }l}
s_{j\geq 2}^{(n)}
                  & s_{j-1}^{(n)} + |t_j^{(n)} | &
                     \mbox { last } j \mbox { with }i_j = 0\,,  \\
                  & s_{j-1}^{(n)} + |t_j| &
                  \mbox { other } j \mbox { with }i_j=0 \,,  \\
                  & s_{j-1}^{(n)} + \frac{1}{n} |t_j| & i_j \not= 0 \,.
\end{array}
\end{array}
\end{equation}
Further, let
\begin{equation}
\begin{array}{l@{\quad }l@{\quad  }l}
u^{(n)}(\tau ) & = n \cdot {\rm sgn} (t_j) e_{i_j}  & \mbox{if }
s_{j-1}^{(n)} \leq  \tau \leq  s_j^{(n)}
\, \&  \, i_j \not= 0\,, \\
& = 0 &   \mbox {if }s_{j-1}^{(n)} \leq  \tau \leq  s_j^{(n)}
              \, \& \,  i_j = 0 \,, \\
           & = 0 & \mbox {if }  \tau \geq s_k^{(n)}\,,
\end{array}
\end{equation}
where $e_1, \ldots , e_r$ are unit vectors in ${\mathbb R}^r$. The
solution $\beta_t^{(n)}$ of the system (\ref{equ431}) associated
with the control $u^{(n)}(\cdot )$ may be written
\begin{equation}
\beta_{s_k^{(n)}}^{(n)} = \beta_{|t_k|}^{n,i_k} \cdots
\beta_{|t_1|}^{n,i_1}  \in  {\hat R}_t(\eta )\,,
\end{equation}
where $\beta_{|\tau|}^{n,i_j} $ is the integral curve of $W_0$ if
$i_j =0$, or the integral curve of $W_0 + n \cdot {\rm
sgn}(\tau)W_{i_j}$ if $i_j \not= 0$, i.e.,
\begin{equation}
\begin{array}{l@{\quad = \quad }l@{\quad \mbox {if }\quad }l}
\beta_{|\tau|}^{n,i_j} & (W_0)_\tau & i_j = 0\,, \\
& (W_0 + n\cdot {\rm sgn}(\tau)W_{i_j})_{\frac{|\tau|}{n}} & i_j \not= 0
\,.
\end{array}
\end{equation}
We note that $(W_0 + n\cdot {\rm sgn}(\tau)W_{i_j})_{\frac{|\tau|}{n}}$
and $(\frac{1}{n}W_0 +{\rm sgn}(\tau)W_{i_j})_{|\tau|}$ describe the
same integral curve on $N$, by virtue of the time-invariance
property of system (\ref{equ431}).  Obviously,
$\beta_{|t_p|}^{n,i_j} \rightarrow \alpha_{t_p}^{i_j}$ as $ n
\rightarrow  \infty $. On the other hand,
\begin{align}
s_k^{(n)} & = \sum_j t_j \cdot 1_{\{ i_j = 0 \} } -\frac{\sum _l
|t_l| \cdot 1_{\{ i_l \not= 0 \} }}{n} + \frac{\sum _l |t_l| \cdot
1_{\{ i_l \not= 0 \} }}{n} \nonumber \\
& =t\,.
\end{align}
Thus, as $n \rightarrow \infty$ we obtain
\begin{equation}
\beta_{s_k^{(n)}}^{(n)}(\eta )
 \rightarrow \alpha_{t_k}^{i_k}
\cdots \alpha_{t_1}^{i_1}(\eta ) = \alpha_t(\eta )\,,
\end{equation}
and hence $\alpha_t(\eta ) \in {\rm cl}\, {{\hat R}_t(\eta )}$.  Because
$\alpha_t(\eta )$ is an arbitrary element in ${\hat R}_t^0(\eta )$,
it follows that $ {\hat R}_t^0(\eta ) \subseteq {\rm cl}\,
{{\hat R}_t(\eta )}$, and since ${\rm cl}\, {{\hat R}_t(\eta )}$ is
closed, it follows in turn that $ {\rm cl}\, {{\hat R}_t^0(\eta )}
\subseteq {\rm cl}\, {{\hat R}_t(\eta )}$.

Next we show ${\rm cl}\, {{\hat R}_t(\eta )} \subseteq {\rm cl}\,
{{\hat R}_t^0(\eta )}$.  Consider $\beta (\eta ) \in {\hat
R}_t(\eta )$ of the form of $\beta_{u_k}^{c_k} \cdot \ldots \cdot
\beta_{u_1}^{c_1}(\eta )$, with $\beta_{u_j}^{c_j} = \exp u_j (W_0
+ c_j^1 W_1 + \ldots  + c_j^r W_r)$ and $c_j = ( c_j^1, \ldots ,
c_j^r)$. Here, $c_j^l$ is the control applied to $W_l$ during time
period $u_j$, so $c_j$ is the control set applied to $W_1, ...
W_r$ during the corresponding time interval $u_j$, with $u_j \in
{\mathbb R}^+$ and $c_j^l \in {\mathbb R}$.  For each
$\beta_{u_j}^{c_j}, \, j =1, \ldots , k$, take $\alpha_j^n $ in
the form
\begin{equation}
\alpha_j^n = \left[\exp \frac{u_j}{n}(c_j^1 W_1) \cdots \exp
\frac{u_j}{n}(c_j^r W_r) \exp \frac{u_j}{n}W_0\right]^n .
\end{equation}
Invoking the Baker-Campbell-Hausdorff formula \cite{Hochschild},
we write
\begin{align}
& \lim_{n \rightarrow \infty }  \alpha_j^n  \\
& = \lim_{n\rightarrow \infty } \left[\exp \frac{u_j}{n}(c_j^1
W_1) \cdots \exp
\frac{u_j}{n}(c_j^rW_r) \cdot \exp \frac{u_j}{n}W_0\right]^n \nonumber \\
 & =  \lim _{n\rightarrow \infty } \exp \left[u_j(W_0
+ c_j^1 W_1 + \cdots  + c_j^r W_r) \right. \nonumber \\
& \quad \left. + \sum_{0 \leq p, q \leq r} \frac{u_j^2}{2n}c_j^p
c_j^q [W_p, W_q] + O\left(\frac{1}{n^2}\right) \right] \nonumber \\
& =  \exp u_j(W_0 + c_j^1 W_1 + \ldots + c_j^r W_r)  =
\beta_{u_j}^{c_j} \,.
\end{align}
Constructing $\alpha_1^n \ldots  \alpha_k^n \in {\hat R}_t^0(\eta
)$ we then obtain
\begin{equation}
\lim _{n\rightarrow \infty } \alpha_k^n \cdots \alpha
_1^n (\eta ) = \beta_{u_k}^{c_k} \cdots \beta_{r_1}^{c_1}(\eta )
=\beta (\eta )\,,
\end{equation}
so that $ \beta (\eta ) \in {\rm cl}\, {{\hat R}_t^0(\eta )}$.  Since
$\beta(\eta )$ is an arbitrary element of ${\hat R}_t(\eta )$, we
arrive at ${\hat R}_t(\eta ) \subseteq {\rm cl}\, {{\hat R}_t^0(\eta )}$
and hence ${\rm cl}\, {{\hat R}_t(\eta )} \subseteq {\rm cl}\,
{{\hat R}_t^0(\eta )}$. We conclude that ${\rm cl}\, {{\hat R}_t(\eta )}
= {\rm cl}\, {{\hat R}_t^0(\eta )}$.

The time $t$ labeling these reachable sets is to be interpreted
as the time interval over which the control operation represented
by $W_0$ is in effect, or ``turned on.''  In fact, $W_0$ is
necessarily {\it always} ``on'' in the augmented system, so the
total time elapsing in the augmented system is the same as the time
interval over which $W_0$ is turned on; hence the reachable sets
${\hat R}_t$ corresponding to these two times are identical.  Of
course, the same coincidence does not hold for the auxiliary system.
However, this is immaterial, since the auxiliary system was only
introduced to exploit the key relationship (\ref{equlem41}). Further,
we may observe that the reachable set ${\hat R}_t^0(\eta )$ of system
(\ref{equ432}), with the control $u(t)= (u_0(t), \ldots , u_r(t))$
assuming values $(e_0, \pm e_1, \ldots , \pm e_r)$, is the same as
the corresponding set for which the control $u(t)$ assumes the
values $e_0, \pm ce_1, \ldots , \pm ce_r$, with $c \in {\mathbb R}^+$.

Since we can take advantage of the result (\ref{equlem41}) in
this manner, it is clearly preferable to study the properties of
${\hat R}_t^0(\eta )$.  The auxiliary system is easier to control, and
the state at time $t$ can be expressed as a composition of
integral curves of $W_i$ in the same style as Eq.~(\ref{equexpre1}).
To do so, let the set of subscripts $j$ with $i_j = 0$ be written
as $\{ p, \ldots, q, s \}$ in increasing order, of course with
$t_p + \ldots + t_q + t_s = t$.  Then we have
\begin{eqnarray}
\alpha_t  & = & (\alpha_{t_k}^{i_k} \cdots
\alpha_{t_{s+1}}^{i_{s+1}}) \cdot (\alpha_{t_s}^0 \cdot
\alpha_{t_{s-1}}^{i_{s-1}} \cdot \alpha_{-t_s}^0)
 \cdot (\alpha_{t_s}^0 \cdot \alpha_{t_{s-2}}^{i_{s-2}}   \nonumber \\
 & &  \cdot
\alpha_{-t_s}^0) \cdots (\alpha_{t_s + t_q}^0 \cdot
\alpha_{t_{q-1} }^{i_{q-1}} \cdot \alpha_{-(t_s + t_q)}^0)  \nonumber \\
& & \cdot (\alpha_{t_s + t_q}^0 \cdot \alpha_{t_{q-2} }^{i_{q-2}}
\cdot \alpha_{-(t_s + t_q)}^0) \cdots \nonumber \\
& & \cdot (\alpha_{t_s + t_q + \cdots + t_p}^0 \cdot
\alpha_{t_{p-1} }^{i_{p-1}} \cdot \alpha_{-(t_s +
t_q + \cdots + t_p)}^0) \cdots  \nonumber \\
& & \cdot (\alpha_{t_s + t_q + \ldots + t_p}^0 \cdot
\alpha_{t_1}^{i_1}
\cdot \alpha_{-(t_s + t_q + \ldots + t_p)}^0) \cdot \alpha_t^0 \nonumber \\
& = &  \beta_0 (\alpha_{t_k}^{i_k}) \cdots \beta_0
(\alpha_{t_{s+1}}^{i_{s+1}}) \cdot
  \beta_{t_s} (\alpha_{t_{s-1}}^{i_{s-1}})
  \cdot \beta_{t_s} (\alpha_{t_{s-2}}^{i_{s-2}}) \cdots  \nonumber \\
& &   \beta_{t_s + t_q} (\alpha_{t_{q-1}}^{i_{q-1}}) \cdot
\beta_{t_s + t_q} (\alpha_{t_{q-2}}^{i_{q-2}})  \cdots \beta_t
(\alpha_{t_{p-1}}^{i_{p-1}}) \cdots \nonumber  \\
& & \cdot \beta_t (\alpha_{t_1}^{i_1}) \cdot
    \alpha_t^0,
\end{eqnarray}
where $\beta _t(\gamma ) = \alpha _t^0 \cdot \gamma \cdot \alpha
_{-t}^0$.  This analysis stimulates us to define the following
three sets of diffeomorphisms:
\begin{align*}
{\rm Exp}\, {\hat{\cal B}}= & \mbox{ the group generated by }
\alpha_t^l, \, t \in {\mathbb R}\,,    \, l=1, \ldots , r, & \\
& \mbox{where } \alpha_t^l \mbox{ is the integral curve of vector field } W_l\,, &\\
F_t= & \cup_{k=1}^\infty  \{ \beta _{t_k}(\gamma_k)\cdot \ldots
\cdot \beta _{t_1}(\gamma_1)| \, \gamma_j \in {\rm Exp}\,
{\hat{\cal B}}, & \\
& 0 \leq t_k \leq \ldots \leq t_1 = t\}\,,& \\
G_t= & \cup_{k=1}^\infty  \{ \beta _{t_k}(\gamma_k)\cdot \ldots
\cdot \beta _{t_1}(\gamma_1)| \, \gamma_j \in {\rm Exp}\,
{\hat{\cal B}}, & \\
& \, \min_j t_j \geq 0, \, \max_j t_j = t\}.&
\end{align*}

By construction,
\begin{equation}
{\hat R}_t^0(\eta )= F_t \alpha_t^0 (\eta )\label{equ433} \,.
\end{equation}
We observe that $F_t$ is a semi-group of diffeomorphisms
included in the the group $G_t$, whose properties are established
in the following lemma.
\begin{lemma}
First, the set $G_t$ is a group.  Furthermore, if $\dim{\hat{\cal
C}}(\eta ) = n-1 = m$ holds for all $\eta \in N$, then $\{ \alpha
(\eta )|  \alpha \in G_t \} = \alpha_t^0 (I(\alpha_{-t}^0(\eta
)))$ is true for all $\eta$, where $I(\nu)$ is the maximal
connected integral manifold containing $\nu \in N$, whose
associated Lie algebra is $\hat{\cal C}$. \label{lem42}
\end{lemma}

\noindent {\bf Proof:}  For $\alpha_1, \alpha_2 \in G_t$, it
is easily seen that $\alpha_1 \cdot \alpha_2 \in G_t$.  Writing
$\alpha \in G_t$ as $\alpha = \beta_{t_k}(\gamma_k)
\cdot \ldots \cdot \beta_{t_1}(\gamma_1)$, we also see that
$\alpha ^{-1} = \beta_{t_1}(\gamma_1^{-1}) \cdot \ldots \cdot \beta_{t_k}
(\gamma_k^{-1})$.  Therefore $G_t$ is a group.

Now, denote the set $\{ \alpha (\eta )| \alpha  \in G_t \}$ by
$B_t(\eta )$.  It is straightforward to show that (i) $B_t(\eta )
= B_t (\xi )$ if $\xi  \in B_t(\eta )$ and (ii) $B_t(\eta ) \cap
B_t(\xi ) = \emptyset  $ if $\xi \not\in B_t(\eta )$
\cite{Kunita_1}.  We can demonstrate that (iii) $\eta \in {\rm
int}\, B_t(\eta)$ under the topology of  $\alpha_t^0
(I(\alpha_{-t}^0(\eta)))$ as follows. By definition, ${\hat
R}_t^0(\eta )$ is the reachable set for the system (\ref{equ432}).
By the same reasoning that leads to Eq.~(\ref{equ433}), we have
${\hat R}_t^0(\alpha_{-t}^0(\eta )) \subset B_t(\eta )$ because
${\hat R}_t^0(\alpha_{-t}^0(\eta )) = F_t \cdot \alpha_t^0 \cdot
\alpha_{-t}^0 (\eta )$.  Since ${\hat R}_t^0(\alpha_{-t}^0(\eta
))$ has a nonempty interior with respect to the topology of
$\alpha_t^0(I(\alpha_{-t}^0(\eta )))$ by Theorem \ref{the42}, we
see that $B_t(\eta )$ contains a non-null open set $U$.  Given $
\mu \in U$, choose $\alpha \in G_t$ such that $\alpha (\eta ) =
\mu $.  Since $\alpha $ is a continuous map, $\alpha ^{-1}(U)$ is
an open set containing $\eta $.

In fact, $\alpha ^{-1}(U)$ is included in $B_t(\eta )$.  We know
that $G_t$ is a group, so $\alpha ^{-1} \in G_t$ if $\alpha \in G_t$.
Letting $ \zeta \in \alpha ^{-1}(U)$, we can find $\chi \in U$, such
that $\chi = \alpha (\zeta) \in U \subset B_t(\eta )$
and also $\chi \in B_t(\zeta)$.  By properties (i) and (ii), we
obtain $\chi \in B_t(\zeta) \cap B_t(\eta ) \neq \emptyset $.
Hence $B_t(\zeta) = B_t(\eta )$ and $ \zeta \in B_t(\eta )$.
Accordingly, $\alpha ^{-1}(U) \subset B_t(\eta )$ and
$\eta \in  {\rm int}\, B_t(\eta)$ under the topology of
$ \alpha_t^0 (I(\alpha_{-t}^0(\eta)))$.

The properties (i)-(iii) imply that $B_t(\eta )$ is maximally
connected and open under the topology of $ \alpha_t^0
(I(\alpha_{-t}^0(\eta )))$.  Thus we have $B_t(\eta ) = \alpha_t^0
(I(\alpha_{-t}^0(\eta )))$ for all $ t > 0$ and $ \eta \in N$. In
addition, it is seen that $B_t(\eta )= \alpha_t^0 (I(\alpha
_{-t}^0(\eta )))= \left ( \begin{array}{c} t_0 \\ M \cap
{\cal D}_A \end{array} \right )$.  The proof of Lemma \ref{lem42} is
now complete.

Based on Lemmas \ref{lem41} and \ref{lem42}, we could
conclude that ${\rm cl}\, {{\hat R}_t(\alpha_{-t}^0(\eta ))} =
{\alpha_t^0(I(\alpha_{-t}^0(\eta )))}$ if we could establish
that $F_t = G_t$.  The following proof takes a slightly different path.
Let ${\rm Exp}\, \widetilde {\hat{\cal B}}$ denote the group of diffeomorphisms
generated by all one parameter groups of transformations with
respect to vector fields belonging to $\widetilde {\hat{\cal B}}$. The sets
$\widetilde {F_t}$ and $\widetilde {G_t}$ are defined in the same
way as $F_t$ and $G_t$, i.e.\ via Eq.~(17), but with
${\rm Exp}\, \widetilde {\hat{\cal B}}$ entering in place of
${\rm Exp}\, {\hat{\cal B}}$.

Obviously, $F_t \subset \widetilde {F_t}$ and $G_t
\subset \widetilde {G_t}$ hold.  We shall now establish
that $\widetilde {F_t} = \widetilde {G_t}$.

\begin{lemma}
Let $X$ be a complete vector field belonging to $\widetilde{\hat{\cal B}}$,
and let $\gamma_t $ be the one-parameter group of transformations
generated by $X$.  Assume $[ {\hat{\cal B}} , {\hat{\cal C}} ](\eta ) \subset
{\hat{\cal B}}(\eta )$ is satisfied for all $\eta $. Then
$d\beta_s (\gamma_t)$ is an isomorphism between ${\hat{\cal B}}(\eta )$
and ${\hat{\cal B}} (\beta_s (\gamma_t)(\eta ))$ for each $\eta $, and
$\widetilde {F_t} = \widetilde {G_t}$ is true for all $ t > 0$.
\label{lem43}
\end{lemma}

\noindent {\bf Proof: }   Since $ \beta _s(\gamma _{t_1}) \cdot
\beta _s(\gamma _{t_2}) =  \beta _s(\gamma _{t_1 +t_2})$ holds, we
have $d\beta _s(\gamma _{t_1 + t_2}) = d\beta _s (\gamma _{t_1})
\cdot d\beta _s(\gamma _{t_2})$. Hence it is enough to prove the
lemma's assertion for sufficiently small $|t|$.  Let $Y_{t,s} =
d \beta _s (\gamma _t) Z$, where $ Z \in \tilde {\hat{\cal B}}$. For
each value of $s$, $\beta _s(\gamma _t)$ with $t \in R$ is
the one parameter group of transformations generated by
$d\alpha _s^0 X$, while
\begin{equation}
\frac{\partial Y_{t,s}}{\partial t}
= -d\beta _s (\gamma _t) [d\alpha _s^0 X, Z ]
= d\beta _s (\gamma _t) [Z, d\alpha _s^0 X] \,.
\end{equation}
Therefore $[ Z, d\alpha _s^0 X ] \in \widetilde{\hat{\cal B}}$ by assumption,
because $ d\alpha_s^0 X$ belongs to
$\widetilde {\hat{\cal C}} = \{ {\hat{\cal C}}(\eta )| \eta  \in N \}$
\cite {Ichihara_1,Ichihara_2}.

Now we fix a point $\eta $ of $N$ and a value of $s \in R$.  Let $
Z^1, \ldots , Z^n$ provide a basis of $\hat{\cal B}$ in an open
neighborhood $U$ of $\eta $.  Then there exist $C^\infty $
functions $f_{ij}$ on $U$ such that $[ Z^i, d\alpha _s^0 X ] =
\sum _{j=1}^n f_{ij} Z^j$ holds in $U$.  Let $\epsilon $ be a
positive number such that $\beta _s (\gamma _t) (\eta ) \in U$ for
$|t| < \epsilon $, noting that $\beta _s(\gamma _t) $ is a
continuous map of $t$ and $\beta _s (\gamma _0)(\eta ) = \eta $.
Then $d\beta _s (\gamma _t )[ Z^i, d\alpha _s^0 X ] = \sum
_{j=1}^n f_{ij} d\beta _s (\gamma _t) Z^j$ for $|t| < \epsilon $.
Set $ V^j(t) = d\beta _s(\gamma _t) Z^j$. Then $V^j(t)$, with
$|t| < \epsilon $, satisfies the linear differential equation
\begin{equation}
\frac{dV^j (t)}{dt} = \sum _{j=1}^n f_{jk} V^k(t) \, \,  j= 1, \ldots , n\,.
\end{equation}
The solution $V^j(t)$ can be written as $V^j(t) = \sum_{k=1}^n
g_{jk}(t) V^k(0)$, where $(g_{jk})$ is a  regular matrix.  Also,
we have $V^k (0) \in \hat{\cal B}(\eta )$ and $V^k(t) \in
{\hat{\cal B}} (\beta_s (\gamma_t))(\eta )$. The map $ d\beta_s
(\gamma_t): {\hat{\cal B}}(\eta ) \rightarrow {\hat{\cal B}}
(\beta_s (\gamma_t))(\eta)$ is bijective because $(g_{jk})$ is a
regular matrix.  Moreover, $d\beta_s(\gamma_t)$ retains the
structure of the Lie bracket with respect to $d\alpha _s^0 X$.
This establishes that $d\beta_s (\gamma_t)$ is an isomorphism
between ${\hat{\cal B}}(\eta )$ and ${\hat{\cal B}}(\beta_s
(\gamma_t))(\eta )$ for $|t| < \epsilon $.  Since $\gamma_t^\prime
\equiv \beta_s (\alpha ) \cdot \gamma_t \cdot \beta_s (\alpha
)^{-1}$ (with $s$ fixed) is a one-parameter group of
transformations generated by $d\beta_s (\alpha ) X$ and $ d\beta_s
(\alpha ) X$ belongs to $\widetilde {\hat{\cal B}}$, we know
$\gamma_t^\prime$ (with $t \in R$) belongs to ${\rm Exp}\,
\widetilde {\hat{\cal B}}$. But ${\rm Exp}\, \widetilde{\hat{\cal
B}}$ is generated by all such $\gamma_t$, so we arrive at the
relationship
\begin{equation}
\beta_t (\alpha )({\rm Exp}\,{\widetilde{\hat{\cal B}}})
\beta_t (\alpha )^{-1} \subset {\rm Exp}\, {\widetilde{\hat{\cal B}}},
\quad \quad \mbox {for } \alpha  \in\widetilde{\hat{\cal B}}\,.
\label{equ434}
\end{equation}

Let $\alpha $ be any element of $\widetilde {G_t}$, written as
\begin{equation}
\alpha =  \beta _{t_k} (\gamma_k) \cdot \ldots \cdot \beta _{t_1}
(\gamma _1), \, \, t_l \geq 0, \, \max_l t_l = t\,.
\end{equation}
By induction we can prove that there exist $\tilde \gamma _k,
\ldots , \tilde \gamma _1$ of ${\rm Exp} \,\widetilde{\hat{\cal B}}$
and $ 0 \leq s_k \leq \ldots \leq s_1 = t $ such that
\begin{equation}
\beta _{t_k} (\gamma _k) \cdot \ldots \cdot \beta _{t_1} (\gamma _1) =
\beta _{s_k} (\tilde \gamma _k) \cdot \ldots \cdot \beta _{s_1} (\tilde
\gamma_1).
\label{equ435}
\end{equation}
Here we only consider the case $k=2$. If $t_2 \leq t_1 $, there is no
need for proof.  Suppose $ t_2 > t_1 $, and set $ t_3 = t_2 -
t_1$.  Then we may write $\beta _{t_2} (\gamma _2) \cdot \beta
_{t_1} (\gamma _1) = \beta _{t_1} (\beta _{t_3} (\gamma _2) \cdot
\gamma _1)$. By relationship (\ref{equ434}), there exists $\tilde
\gamma _1$ of ${\rm Exp}\, \widetilde{\cal B}$ such that $\beta _{t_3}
(\gamma _2) \cdot \gamma _1 \cdot \beta _{t_3} (\gamma _2)^{-1} =
\tilde \gamma _1$, i.e., $ \beta _{t_3} (\gamma _2) \cdot \gamma
_1 = \tilde \gamma _1 \cdot \beta _{t_3} (\gamma _2)$.  This
implies
\begin{equation}
\begin{split}
\beta _{t_2} (\gamma _2) \cdot \beta _{t_1}(\gamma _1) & = \beta
_{t_1} (\beta _{t_3} (\gamma _2) \cdot \gamma _1 ) = \beta_{t_1}\,
( \tilde \gamma _1 \cdot \beta _{t_3} (\gamma _2) ) \\
& = \beta _{t_1} (\tilde \gamma _1) \cdot \beta _{t_2} (\gamma
_2)\,.
\end{split}
\end{equation}
More detailed proofs may be found in Refs.~\cite {Kunita_1,Lan_2}.

\begin{theorem}
Suppose that $\dim \,{\hat{\cal C}}(\eta ) = n-1 = m $ holds for
all $\eta \in N$, and suppose that $[ {\hat{\cal B}}, {\hat{\cal
C}} ] (\eta ) \subset  {\hat{\cal B}} (\eta )$ holds for all $\eta
$. Let $I(\eta )$ be the maximally connected integral manifold
containing $\eta $ whose corresponding Lie algebra is $\hat{\cal
C}$.  Then $\alpha_t^0 ( I (\eta )) = {\hat R}_t (\eta )$.
\label{the44}
\end{theorem}

\noindent {\bf Proof: } Clearly we have $\{ \alpha  \alpha_t^0(\eta )|
\alpha  \in F_t \}  \subset \{ \alpha  \alpha _t^0(\eta )|  \alpha
\in \widetilde{F_t}\}$.  In fact, the closures of these two sets
coincide.  Since $\widetilde {F_t} = \widetilde {G_t} \supset  G_t$,
it is seen that
\begin{equation}
\begin{array}{l@{\quad = \quad }ll}
{\rm cl}\, {{\hat R}_t^0 (\eta )} & {\rm cl}\,{ \{\alpha \alpha_t^0(\eta )|
\alpha \in F_t \}} & \\
&{\rm cl}\,{\{\alpha \alpha_t^0(\eta )|\alpha \in \widetilde{F_t}\}}  & \\
& {\rm cl}\,{\{\alpha \alpha_t^0(\eta)|\alpha \in \widetilde {G_t}\}}
& \mbox {(by Lemma \ref{lem43})} \\
& {\rm cl}\,{\alpha_t^0 (I(\alpha_{-t}^0(\alpha_t^0 (\eta ))))}&
\mbox{(by Lemma \ref{lem42})} \\
& {\rm cl}\,{\alpha_t^0 (I(\eta ))}\,. &
\end{array}
\end{equation}

But Lemma \ref{lem41} tells us that $ {\rm cl}\,{{\hat R}_t^0
(\eta )} = {\rm cl}\, {{\hat R}_t (\eta )} $, so we obtain ${\rm
cl}\, {{\hat R}_t (\eta )} = {\rm cl}\, { \alpha_t^0 ( I (\eta
))}$. And from Lemma \ref{lemthe43} we know that $ {\rm int}\,
{\hat R}_t (\eta ) = {\rm int}({\rm cl}\, {\hat R}_t (\eta ))$,
which implies ${\rm int}\, {\hat R}_t(\eta ) = \alpha_t^0 (I
(\eta))$ under the topology of $\alpha_t^0 (I (\eta ))$. Finally,
${\hat R}_t( \eta ) \subset \alpha_t^0 (I(\eta ))$ by Theorem
\ref{the42}, and we arrive at ${\hat R}_t (\eta ) = \alpha_t^0
(I(\eta ))$.

\subsection{Strong Analytic Controllability of the Actual System}

In subsection 4.2, we investigated the reachable set at time $t$
of the time-independent augmented system formed by enlarging the
state space to include an extra dimension corresponding to the
variable $t$.  Now we return to the original quantum control system
(\ref{equ421}) to discover conditions under which it is
strongly analytically controllable.

\begin{theorem}
For the control system defined by Eq.~(\ref{equ421}), let
\begin{equation}
\begin{array}{l}
{\cal B}(t) = {\cal L} (H_1 (t), \ldots , H_r (t)) \\ \\
B_1 = - [ H_0, {\cal B} ] + \frac{\partial }{\partial t}
{\cal B} \\ \vdots \\
B_n = - [ H_0, B_{n-1} ] + \frac{\partial }{\partial t} B_{n-1} \\
\vdots  \\ {\cal C} = {\cal L} \{ {\cal B}, B_1,
\ldots , B_n, \ldots \} \,.
\end{array}
\end{equation}
Suppose $\dim \,{\cal C} (t) \psi (t) = m$ holds
for all $\psi  \in  M \cap {\cal D}_A $, and $[{\cal B},
{\cal C}](t) \subset  {\cal B}(t)$ is the case
for all $t$. Then the time-dependent quantum control system
(\ref{equ421}) is strongly analytically controllable.
\label{the45}
\end{theorem}

\noindent {\bf Proof: }  We apply Theorem \ref{the44} to the
augmented control system (\ref{equ431}).  To do so, we need
to examine the Lie algebras $\cal B$ and $\cal C$ for this problem.
For $\cal B$ we readily find
\begin{equation}
\begin{split}
{\cal B} = & {\cal L} \{ W_1, \ldots , W_r \} \vspace{0.5cm}\\
 = & {\cal L} \left \{  \left (  \begin{array}{c}
                                   0 \\
                                   H_1(t)
                                   \end{array}  \right ) ,
                                   \ldots ,
                          \left (  \begin{array}{c}
                                   0 \\
                                   H_r(t)
                                   \end{array}\right)
   \right  \} \psi (t) \\
 =  & \left ( \begin{array}{c}
              0 \\
              {\cal L} \{ H_1(t), \ldots , H_r(t) \}
              \end{array}
              \right ) \psi (t)
= \left ( \begin{array}{c}
              0 \\
              {\cal B} (t) \psi (t)
              \end{array}
              \right ).
\end{split}
\end{equation}

Next let us construct $\cal C$.  For any
\begin{equation}
W(\eta ) = W(t, \psi ) = \left ( \begin{array}{c} 0 \\
              H(t) \psi (t) \end{array} \right )  \in {\cal B}\,,
\end{equation}
where $\eta \in N$, we have
\begin{equation}
\begin{split}
& {\rm ad}_{W_0} W  = [W_0, W ]  = \left [ \left (
\begin{array}{c}
                       1 \\
                       H_0(t) \psi (t)
                       \end{array}  \right ),
               \left ( \begin{array}{c}
                       0 \\
                       H(t) \psi (t)
                       \end{array}  \right ) \right ]  \\
& =  \frac{ \partial \left ( \begin{array}{c}
                       0 \\
                       H(t) \psi (t)
                       \end{array}  \right ) }{\partial (t,\psi )}
                     \left ( \begin{array}{c}
                       1 \\
                       H_0(t) \psi (t)
                       \end{array}  \right )
    -\frac{ \partial \left ( \begin{array}{c}
                       1 \\
                       H_0(t) \psi (t)
                       \end{array}  \right ) }{\partial (t,\psi)} \\
  & \cdot                   \left ( \begin{array}{c}
                       0 \\
                       H(t) \psi (t)
                       \end{array}  \right )
= \left \{ \begin{array}{c}
                0 \\
                -[H_0, H ] + {\partial H}/{\partial t}
                \end{array} \right \} \psi (t) \,.
\end{split}
\end{equation}
Similarly,
\begin{equation}
{\rm ad}_{W_0} {\cal B} = \left ( \begin{array}{c} 0  \\
-[ H_0 , {\cal B} ]
+ {\partial {\cal B}}/{\partial t}
\end{array} \right ) \psi (t)\,.
\end{equation}
Setting $B_1 = -[H_0, {\cal B}] + \partial {\cal B}/\partial t$, we may
then derive
\begin{equation}
\begin{split}
& {\rm ad}_{W_0}^2 {\cal B}  = {\rm ad}_{W_0} {\rm ad}_{W_0} {\cal B}  \\
& = {\rm ad}_{W_0}
\left ( \begin{array}{c} 0 \\
         B_1 \psi (t)
        \end{array} \right) = \left ( \begin{array}{c} 0 \\
          -[ H_0, B_1 ] + {\partial B_1}/{\partial t}
          \end{array} \right ) \psi (t)\,.
\end{split}
\end{equation}
Continuing in this fashion with
\begin{equation}
B_n = -[H_0, B_{n-1}] + {\partial B_{n-1}}/{\partial t}
\end{equation}
for $n=2,3, \ldots $, we find
\begin{equation}
\begin{array}{ll}
{\rm ad}_{W_0}^n {\cal B} & = \left ( \begin{array}{c}
                               0 \\
                -[H_0, B_{n-1}] + \frac{\partial B_{n-1}}{\partial t}
                               \end{array} \right ) \psi (t)  \\
                     & = \left ( \begin{array}{c}
                               0 \\
                              B_n \psi (t)
                               \end{array} \right ) \,.
\end{array}
\end{equation}
Thus
\begin{equation}
\begin{split}
{\cal C} & = {\cal L} \{ {\cal B}, {\rm ad}_{W_0} {\cal B}, \ldots ,
{\rm ad}_{W_0}^n {\cal B}, \ldots \} \\
& = {\cal L} \left \{ \left ( \begin{array}{c}
                                 0 \\
                                 {\cal B} (t) \psi (t)
                                 \end{array}  \right ),
                         \ldots ,
                        \left ( \begin{array}{c}
                                 0 \\
                                 B_n (t) \psi (t)
                                 \end{array}  \right ), \ldots  \right \} \\
& =  \left ( \begin{array}{c}
              0 \\
              {\cal L} \{ {\cal B}(t), B_1(t), \ldots , B_n(t), \ldots \} \psi (t)
              \end{array} \right ) \\
& =  \left ( \begin{array}{c}
              0 \\
              {\cal C}(t) \psi (t)
              \end{array} \right )\,.
\end{split}
\end{equation}

From the assumption  that $[ {\cal B},{\cal C}](t)
\subset  {\cal B}(t)$, $ \forall (t)$, we have
\begin{equation}
[{\cal B} , {\cal C}](t) \psi (t) \subset
{\cal B}(t)\psi (t), \, \forall (t) \,.
\end{equation}
Hence
\begin{equation}
\left [ \left ( \begin{array}{c}
                      0 \\
                      {\cal B} \psi
                      \end{array} \right ),
              \left ( \begin{array}{c}
                      0 \\
                      {\cal C} \psi
                      \end{array} \right ) \right ] \subset
              \left ( \begin{array}{c}
                      0 \\
                      {\cal B} \psi
                      \end{array} \right )\,,
\end{equation}
so that $[ {\cal B}, {\cal C}] (\eta ) \subset {\cal B}(\eta )$,
$\forall \eta \in N$.

By assumption, $\dim \, {\cal C}(t) \psi(t) = m, \,
\forall \psi \in M \cap {\cal D}_A $, which implies that
$\dim \,{\cal C}(\eta ) = m =n-1$ holds for all $\eta \in N$.
According to Theorem \ref{the44}, $\alpha_t^0(I(\eta )) =
{\hat R}_t(\eta ), \, \forall t > 0$, and since
$\alpha_t^0(I(\alpha_{-t}^0 (\eta ))) = \left (\begin{array}{c}
t_0 \\ M \cap {\cal D}_A  \end{array} \right)$, we obtain
$ \alpha_t^0 (I(\eta )) = \left ( \begin{array}{c} t+t_0 \\
M \cap {\cal D}_A \end{array} \right ) $.

Let $\pi : N \rightarrow M \cap {\cal D}_A $ be the projection map
that in effect annihilates the time-dimension of the augmented problem
corresponding to the variable $t$, and brings us back to the original
control system. In fact, the extension and projection maps mediate
a one-to-one correspondence between the states of the augmented system
and those of the original system.  The simplicity of this relationship
stems from the fact that $t$ is a strictly increasing variable.

To reiterate our strategy: We have dealt with the explicit
time-dependence of the original control problem by adding an extra
dimension to its state space, such that, as viewed in the augmented
space, the augmented control problem is time-independent.
After analyzing controllability within this extension, the results
are projected to the original space by removing the extra time
dimension, recovering the exact states of the original system.

Accordingly, $ \pi (\alpha_t^0 (I(\eta ))) = M \cap {\cal D}_A $,
while $ \pi {\hat R}_t(\eta ) = R_{t+t_0} (\psi ),\,
\forall \psi \in M \cap {\cal D}_A $.   Hence
$R_{t}(\psi ) = M \cap {\cal D}_A , \forall t > t_0$, and the
system (\ref{equ421}) is strongly analytically controllable
on $M$.

We may note that upon introducing the Lie algebra
${\cal A}(t) = {\cal L} \{ H_0 (t), H_1
(t), \ldots ,$  $ H_r (t) \}$, it is readily established
from property (\ref{equnew1}) that  ${\cal B} \subset
{\cal C} \subset {\cal A}$ for all $t$.

To complete the formal analysis, we state two corollaries that
devolve immediately from Theorem \ref{the45}:
\begin{corollary}
From the operators $H_i$ entering control
system (\ref{equ421}), form the Lie algebras
${\cal B}={\cal L}\{ H_1, \ldots , H_r \} $
and ${\cal C}={\cal L}\{{\cal B},
{\rm ad}_{H_0}{\cal B}, \ldots ,
{\rm ad}_{H_0}^n {\cal B}, \ldots \}$.  Suppose that
the $H_i$ do not possess explicit dependence on the time $t$,
that $\dim \,{\cal C} \psi (t) = m$ holds for all
$\psi \in M \cap {\cal D}_A $, and that
$[{\cal B},{\cal C}]\subset{\cal B}$
is satisfied.  Then the time-invariant system (\ref{equ421}) is
strongly analytically controllable.
\end{corollary}

\begin{corollary}
For the control system (\ref{equ421}), form the Lie algebra
${\cal B}(t) = {\cal L}(H_1(t), \ldots , H_r(t))$,
and suppose that $\dim \,{\cal B}(t) \psi (t) = m$
holds for all $\psi \in M \cap {\cal D}_A $.  Then system
\ref{equ421} is strongly analytically controllable.
\end{corollary}

\noindent
The latter corollary follows because $[{\cal B},
{\cal C}](t) \subset {\cal B} (t)$ must hold, once
$\dim \,{\cal B}(t) \psi (t) = m$.

\section{Examples of Strong Analytic Controllability}

In this section, we present three examples that meet the criteria
for analytic controllability enunciated in Theorem 4.7.  The examples
selected are relevant to problems of interest in mathematical physics
or engineering applications of quantum mechanics.

\noindent {\it Example 1 } \quad  The strong analytic controllability
theorem can be applied to the simple degenerate parametric oscillator,
a problem of importance in physics and engineering.  Introducing
an appropriate effective Hamiltonian allows the corresponding control
system to be written in the form \cite{Walls}
\begin{equation}
i \frac{\partial }{\partial t} \psi = \left\{ \omega (t) a^\dagger a
+ \frac{1}{2}\chi (t)\left[e^{-2i\omega t}(a^\dagger )^2
+e^{2i\omega t}a^2\right] \right\} \psi \,. \label{equ54}
\end{equation}
Here $a^\dagger$ and $a$ represent, in turn, the creation and
annihilation operators of the pump mode of frequency $\omega(t)$,
while $\chi (t)$ is the time-dependent coupling function related
to the second-order nonlinear susceptibility of the pumped medium.
We may consider $\omega (t)$ and $\chi (t)$ as control functions
playing the role of the $u_l$ in Eq.~(\ref{equ421}), since they
are real and can be adjusted to piecewise-constant functions of
time $t$, outside the system itself.

Following precedent \cite{Brif,Gerry_1,Gerry_2,Zhang}, we define
the operators
\begin{equation}
K_+= \frac{1}{2} (a^\dagger )^2 \,, \quad
K_-= \frac{1}{2} a^2 \,, \quad
K_0= \frac{1}{2} (a^\dagger a + a a^\dagger ) \,,
\end{equation}
which satisfy the commutation relations of $SU(1,1)$, thus
\begin{equation}
[K_0, K_\pm ] = \pm K_\pm \,, \quad [K_+, K_-] = -2 K_0\,.
\end{equation}

Setting
\begin{eqnarray}
H_0 &=&  -iK_0\,, \\
H_1 &=& -{i \over 2}[e^{-2i\omega t}K_+ +e^{2i\omega t}K_-]\,, \\
H_2&=& {1 \over 2} [e^{-2i\omega t}K_+ -e^{2i\omega t}K_-]/2\,,
\end{eqnarray}
the control system (\ref{equ54}) may be written in the more familiar
form
\begin{equation}
\frac{\partial}{\partial t} \psi =\left[\omega (t)H_0 + \chi(t)
H_1(t)\right] \psi \,.  \label{equ55}
\end{equation}
The skew-Hermitian operators $H_0$, $H_1$, and $H_2$ satisfy
the commutation relations
\begin{equation}
[H_0, H_1]  = -H_2\,, \quad  {[H_0, H_2]} =  H_1\,, \quad {[H_1,
H_2]}= H_0 \,.
\end{equation}

We observe that the system (\ref{equ55}) does not have a drift
term in the usual sense, because the factor $\omega(t)$ can be
manipulated externally.  We also see immediately that ${\cal A} =
{\cal B} = {\cal C}={\cal L}\{H_0,H_1, H_2\}$, and the second
condition of Theorem 4.7 is obviated. In addition, $H_0$ has
eigenvectors $|mk\rangle $, with $m= 0, 1, \ldots $ and $k =1/4,\
3/4$, which span an analytic domain ${\cal D}_A$ \cite{Gerry_1,
Zhang}.  Consequently, we can choose a manifold $M$ such that
$\dim {\cal C} \psi =\dim M$ $\forall \psi \in {\cal D}_A \cap M$.
All conditions of Theorem 4.7 being met, the system (\ref{equ54})
is strongly analytically controllable on $M$.

\noindent {\it Example 2} \quad Defining $Q=i\partial _t +\partial
_{x_1 x_1} +\partial _{x_2 x_2}$, the Schr\"odinger equation for a
free particle moving in two spatial dimensions may be expressed
simply as $Qu=0$.  Determination of the maximal symmetry algebra
of this equation leads to the following set of nine operators,
which form the basis of a nine-dimensional complex Lie algebra:
\cite{Boyer}
\begin{equation}
\begin{split}
& K_2= -t^2 \partial _t -t(x_1\partial _{x_1} +x_2 \partial
_{x_2}) -t+(i/4)(x_1^2+x_2^2)\,, \\
& K_{-2}=\partial _t,  \quad J=x_1\partial _{x_2}-x_2\partial _{x_1}, \quad B_j =-t\partial _{x_j}+ix_j/2, \\
&  P_j =\partial _{x_j}, \quad E=i, \quad D =x_1 \partial
_{x_1}+x_2\partial _{x_2}+2t\partial _t+1,
\end{split}
\end{equation}
with $j=1,2$.  Of immediate concern is the real Lie algebra
spanned by this basis, i.e., the Schr\"odinger algebra, which has,
as alternative basis, the operators $B_j$, $P_j$, and $E$ (yielding
the five-dimensional Weyl algebra), plus the operator
$J$ and the three operators defined by $L_1=D$, $L_2=K_2+K_{-2}$,
and $L_3=K_{-2}-K_2$.  The pertinent nonvanishing commutators
are specified by \cite{Boyer}:
\begin{equation}
\begin{split}
& [L_1, L_2]=-2L_3\,, ~ [L_3, L_1]=2L_2\,, ~ [L_2, L_3]=2L_1\,, \\
& [L_1,B_j]=B_j\,, ~ [L_1, P_j]=-P_j\,, [P_j, J]=(-1)^{j+1}P_l\,,\\
& [B_j, J]=(-1)^{j+1}B_l\,, ~ [L_2, B_j]=-P_j\,,
~ [L_3, B_j]= -P_j\,, \\
& [L_2, P_j]=B_j\,, ~ [L_3, P_j]=-B_j,\, [P_j, B_j]=E/2\,,
\label{equ56}
\end{split}
\end{equation}
where $j,l=1,2, j \neq l$.

Now we consider the controllability of the system
\begin{equation}
\frac{\partial }{\partial t} \psi =
[L_2+u_1(t)L_1+u_2(t)L_3+u_3(t)P_1+u_4(t)J]\psi \,. \label{equ510}
\end{equation}
In this case there is a time-dependent drift term in the vector
field driving $\psi$.
The relations (\ref{equ56}) imply the equalities ${\cal B}=
{\cal C} ={\cal L}\{ L_1,L_2, L_3, P_1, P_2, B_1, B_2, J, E\}$,
while the required analytic domain $\cal D_A$ is furnished by the span
of the eigenfunctions $\psi _{n, m}$ of $L_3$.  These take the
explicit, time-dependent form \cite{Boyer}
\begin{equation}
\begin{split}
 \psi _{n, m}& = (2^{m+n+1}\pi n!m!)^{-1/2} \exp [i\pi (m+n-1)/2]\\
& \quad \times \exp \left [ \frac{(v_1^2+v_2^2)(1-iv_3)}{4}
\right] \left ( \frac{v_3+i}{v_3-i} \right )^{(m+n)/2}\\
& \quad \times \frac{H_m(v_1/\sqrt{2})H_n(v_2/\sqrt{2})}{v_3-i}\,,
\end{split}
\end{equation}
where $x_1=v_1(1+v_3^2)^{1/2}$, $x_2=v_2(1+v_3^2)^{1/2}$, and
$t=v_3$. It follows as before that the system (\ref{equ510}) is
strongly analytically controllable.

\noindent {\it Example 3} \quad  A quantum control system with
position-dependent effective mass $m= (2Ax)^{-1}$ has been described
by the time-dependent Schr\"odinger equation \cite{ZhangShengli}
\begin{equation}
i\frac{\partial }{\partial t} \psi =
[iBI_0+u_1(t)A(t)I_0I_-+iu_2(t)C]\psi\,, \label{equ57}
\end{equation}
where $B,\, C \in {\mathbb R}$ and $A(t)$ is a real function of
time $t$ but in general not piecewise-constant.  The operators
$I_0$ and $I_\pm$, which are independent of time, provide a basis
for an $su(1,1)$ algebra, and have the concrete realization
\begin{equation}
 I_-=-\partial_x\,, \quad
 I_0=x\partial_x+1\,, \quad I_+=x^2\partial_x+2x\,,
\end{equation}
which satisfies the commutative relations
\begin{equation}
[I_0, I_\pm]=\pm I_\pm, \quad [I_-, I_+]=-2I_0 \,. \label{equ58}
\end{equation}
This effective-mass problem arises in the study of semiconductor
heterostructures and, more generally, of inhomogeneous crystals
\cite{Leblond}.  In the semiconductor application, the effective
mass of a carrier depends spatially on the graded composition of
the semiconductor alloys used in the barrier and well regions
of the microstructures \cite {Chetouani}.

The wave functions of the stationary states of Eq.~(\ref{equ57})
can be written as
\begin{equation}
\begin{split}
& \psi _E(t,x) = \frac{1}{\sqrt{2\pi }} \exp \left \{-iE \int_0^t
B(\sigma )d\sigma  + \int_0^t[-C(\sigma ) \right. \\
& \left. - \frac{1}{2} B(\sigma )]d\sigma \right \} \times \exp
\left \{ -a_1(t) \left ( x\partial _{xx}+
\partial _x \right ) \right \} x^{-iE-1/2} \\
& =\frac{1}{\sqrt{2\pi }} \exp \left \{-iE \int_0^t B(\sigma )
d\sigma  + \int_0^t[-C(\sigma )- \frac{1}{2} B(\sigma
)]d\sigma \right \} \\
& \quad \sum_{n=0}^{\infty }
\prod_{l=0}^n(iB(t)E+\frac{1}{2}+l)^2[-a_1(t)]^n  \times
\frac{x^{-iE-n-1/2}}{n!}\,.
\end{split}
\end{equation}
These eigenfunctions span the analytic domain relevant to
Theorem 4.7.

Let us define
\begin{equation}
H_0=BI_0+u_2(t)C,, \quad H_1= -iA(t)I_0I_-, \,
\end{equation}
where we take $u_2(t)=-B/{2C}$.
Eq.~(\ref{equ57}) can be recast as the control
system
\begin{equation}
\frac{\partial }{\partial t} \psi = [H_0 + u_1(t) H_1]\psi \,.
\end{equation}
Here the drift term is time-independent. Using the commutation
relations (\ref{equ58}), we obtain $[H_0, H_1]= -BH_1$. Obviously,
${\cal B} ={\cal C} \subset {\cal A}$, so $[ {\cal B},  {\cal C}]=
{\cal B}$. Choosing a manifold $M$ such that $\dim M= \dim  {\cal
C} \psi $ for all $\psi \in M$, we are assured that system
(\ref{equ57}) is strongly analytically controllable.

\section{Conclusions}

In this paper, we have formulated the time-dependent quantum control
problem and studied its controllability.  Acknowledging the unbounded 
nature of operators commonly involved in quantum control systems, 
our analysis has been predicated on the existence of an analytic 
domain \cite{Nelson} on which exponentiations of such operators 
are guaranteed to converge.  Within this framework, we have extended 
the established treatment of time-independent quantum control 
problems by introducing an augmented system described in a state 
space that is enlarged by one dimension, yet embodies the true 
dynamics of the original system.  With the aid of techniques and 
results developed by Kunita \cite{Kunita_1,Kunita_2}, we are able 
to explicate the one-dimension-reduced controllability of the augmented 
system.  Projection onto the original state space then yields 
a proof of the analytic controllability of the original time-dependent 
quantum control system, under conditions similar to those 
required in the time-independent case. The theorem so established 
has been illustrated with examples drawn from mathematical 
physics and systems engineering.

\section*{Acknowledgment}

This research was supported in part by the U.~S.\ Army Research
Office (TJT) under Grant W911NF-04-1-0386 and by the U.~S.\
National Science Foundation under Grants DMS01-03838 (QSC) and
PHY-0140316 (JWC).  JWC would also like to acknowledge partial
support from FCT POCTI, FEDER in Portugal and the hospitality of
the Centro de Ci\^encias Mathem\'aticas at the Madeira Math
Encounters.


\begin{thebibliography}{99}

\footnotesize

\bibitem {Kuriksha} A. Kuriksha, {\it Quantum Optics and Optical
Location}, Sovetskoe Radio, 1973.

\bibitem {Bradley} D. J. Bradley, {\it The Laser: the Dynamo of
the Twenty-First Century}, Journal of Russian Society of Arts,
Nov, 1977, pp. 3-20.

\bibitem {Butkovskiy_1} A. G. Butkovskiy and Yu. I. Samoilenko,
{\it Control of Quantum Systems}, Automation and Remote Control,
Vol. 4, April, 1979, pp. 485-502.

\bibitem {Butkovskiy_2} A. G. Butkovskiy and Yu. I. Samoilenko,
{\it Control of Quantum Systems}, Automation and Remote Control,
Vol. 5, May, 1979, pp. 629-645.

\bibitem {Huang_1} Garng M. Huang, T. J. Tarn and John W. Clark,
{\it On the Controllability of Quantum-mechanical Systems}, J.
Math. Phys. 24 (11), 1983, pp. 2608-2618.

\bibitem {Blaquiere_1} A. Blaquiere, {\it Information Complexity
and Control in Quantum Physics}, Proceedings of the 4th International
Seminar on Mathematical Theory of Dynamical Systems and Microphysics,
Udine, 1985, edited by  A. Blaquiere, S. Diner, and G. Lochak.

\bibitem {Blaquiere_2} A. Blaquiere, {\it Modeling and Control of
Systems in Engineering, Quantum Mechanics, Economics and
Biosciences}, Proceedings of the Bellman Continuum Workshop,
Sophia Antipolis, 1988.

\bibitem {Butkovskiy_3} A. G. Butkovskiy and Yu. I. Samoilenko,
{\it Control of Quantum Mechanical Processes and Systems}, Kluwer,
Dordrecht, 1990.

\bibitem {Ezawa} {\it Quantum Control and Measurement},
Proceedings of the ISQM Satellite Workshop ARL, Hitachi, 1992,
edited by H. Ezawa and Y. Murayama.

\bibitem {Gordon} R. Gordon and S. A. Rice, {\it Active Control of
the Dynamics of Atoms and Molecules }, Ann. Rev. Phys. Chem. 48,
1997, pp. 601-641.

\bibitem {Rabitz} H. Rabitz, R. de Vivie-Riedle, M. Motzkus, and
K. Kompa, {\it Whither the Future of Controlling Quantum Phenomena
}, Science, 288, 2000, pp. 824-828.

\bibitem {Lloyd_1} S. Lloyd, {\it Coherent Quantum Feedback},
Physical Review A, Vol. 62, 2000, pp. 022108(1-12).

\bibitem {Lloyd_2} S. Lloyd and S. L. Braunstein, {\it
Quantum Computation over Continuous Variables }, Physical Review
Letters, vol. 82, 1999, pp. 1784-1787.

\bibitem {Brockett} R. W. Brockett, {\it Nonlinear Systems and
Differential Geometry}, Proceedings of  the IEEE, vol. 64, No. 1,
Jan 1976, pp. 61-72.

\bibitem {Slichter} C. P. Slichter, {\it Principles of Magnetic
Resonance}, 3rd ed., Springer-Verlag, New York, 1990.

\bibitem {Ernst_1} C. P. Ernst, G. Bodenhausen, and A. Wokaun,
{\it Principles of Nuclear Magnetic Resonance in One and Two
Dimensions }, Oxford University Press, Oxford, 1987.

\bibitem {Elmsley} L. Elmsley and A. Pines, {\it Lectures on
Pulsed NMR}, 2nd ed., Proceedings of the International School of
Physics " Enrico Fermi", Varenna, 1994.

\bibitem {Lloyd_3} S. Lloyd, {\it Almost Any Quantum Logic Gate is Universal},
Phys. Rev. Lett. 75, 1995, pp. 346-349.

\bibitem {Deutsch} D. Deutsch, A. Barenco, and A. Ekert, {\it Universality
in Quantum Computation}, Proc. Roy. Soc. (London), Ser. A 449,
1995, pp. 669-677.

\bibitem {Akulin} V. M. Akulin, V. Gershkovich, and G. Harel, {\it
Nonholonomic Quantum Devices }, Phys. Rev. A, vol. 64, 2001, pp.
012308(1-8).

\bibitem {Clark} J. W. Clark, {\it Control of Quantum Many-Body Dynamics:
Designing Quantum Scissors}, in {\it Condensed Matter Theories},
Vol. 11, 1996, edited by E. V. Ludena, P. Vashishta, and R. F.
Bishop, Nova Science Publishers, Commack, NY, pp. 3-19.

\bibitem {Ramakrishna_1} V. Ramakrishna, M. V. Salapaka, M.Dahleh,
H. Rabitz and A. Peirce, {\it Controllability of Molecular
Systems}, Phys. Rev. A 51, 1995, pp. 960-966.

\bibitem {Ramakrishna} V. Ramakrishna and H. Rabitz, {\it Relation
Between Quantum Computing and Quantum Controllability}, Phys. Rev.
A, vol. 54, 1996, pp. 1715-1716.

\bibitem {Sussmann}  H\'ector J. Sussmann \& Velimir Jurdjevic,
{\it Controllability of Nonlinear Systems}, Journal of
Differential Equations, 12, 1972, pp. 95-116.

\bibitem{Jurdjevic} V. Jurdjevic and H. J. Sussmann, {\it Control Systems
on Lie Groups}, J. Differ. Equat., Vol. 12, 1972, p. 313-329.

\bibitem {Lloyd_4} S. Lloyd and J.~J.~E. Slotine, {\it Analog Quantum
Error Correction }, Phys.  Rev. Lett. 80, 1998, pp. 4088-4091.

\bibitem {Braunstein_1} S. L. Braunstein, {\it Error Correction.
for Continuous Quantum Variables}, Phys. Rev. Lett. 80, 1998, pp.
4084-4087.

\bibitem {Braunstein_2} S. L. Braunstein, {\it Quantum Error Correction for Communication with Linear
Optics}, Nature (London) vol. 394, 1998, pp. 47-49.

\bibitem {Braunstein_3} S. L. Braunstein and H. J. Kimble,
{\it Teleportation of Continuous Quantum Variables}, Phys. Rev.
Lett. 80, 1998, pp. 869-872.

\bibitem {Furusawa} A. Furusawa {\it et al.}, {\it Unconditional Quantum Teleportation}, Science 282,
1998, pp. 706-709.

\bibitem {Huang_2} Guang M. Huang, {\it  Control of Quantum Systems}, Doctoral Dissertation, Washington University,  1980.

\bibitem {Nelson} Edward Nelson,  {\it Analytic Vectors},  Annals of
Mathematics, Vol. 70, No. 3, Nov., 1959, pp. 572-615.

\bibitem {Sakawa} Y. Sakawa, {\it Feedback Stabilization of Linear
Diffusion Systems}, SIAM J. Control Optim., 21, 1983, pp. 667-676.

\bibitem {Keulen} B. Keulen, {\it Redheffer's Lemma and ${\cal H}_\infty
$-control for Infinite-dimensional Systems}, SIAM J. Control
Optim., 32, 1994, pp. 261-278.

\bibitem {Morris} K. A. Morris, {\it ${\cal H}_\infty $-output
Feedback of Infinite-dimensional Systems via Approximation},
System \& Control Lett., 44, 2001, pp. 211-217.

\bibitem {Morris_2} K. A. Morris, {\it Design of Finite-dimensional
Controllers for Infinite-dimensional Systems by Approximation},
Journal of Mathematical Systems, Estimation, and Control, Vol. 4,
No. 2, 1994, pp. 1-30.

\bibitem {Costa} O. L. V. Costa and C. S. Kubrusly, {\it State
Feedback $H_\infty $-control for Discrete-time
Infinite-dimensional Stochastic Bilinear Systems}, Journal of
Mathematical Systems, Estimation, and Control, Vol. 6, No. 2,
1996, pp. 1-32.

\bibitem {Weiss_2}  G. Weiss and R. Rebarber, {\it Optimizability and Estimatability for Infinite-Dimensional Linear
Systems}, SIAM Journal on Control and Optimization, Vol. 39, 2000,
pp. 1204-1232.

\bibitem {Wen} J. T. Wen, {\it Finite Dimensional Controller
Design for Infinite Dimensional Systems: The Circle Criterion
Approach}, Systems \& Control Letters, Vol. 13, 1989, pp. 445-454.

\bibitem {Corduneanu} C. Corduneanu, {\it Integral Equations and
Stability of Feedback Systems}, Academic Press, New York, 1973.

\bibitem {Leonov} G. A. Leonov, D. V. Ponomarenko, and V. B.
Smirnova, {\it Frequency-Domain Methods for Nonlinear Analysis},
World Scientific, Singapore, 1996.

\bibitem {Logemann} H. Logemann, {\it Circle Criteria, Small-gain
Conditions and Internal Stability for Infinite-dimensional
Systems}, Automatica, Vol. 27, 1991, pp. 677-690.

\bibitem {Wexler} D. Wexler, {\it On Frequency Domain Stability
for Evolution Equations in Hilbert Spaces via the Algebraic
Riccati Equation}, SIAM J. Math. Analysis., Vol. 11, 1980, pp.
969-983.

\bibitem {Rebarber} R. Rebarber and H. Zwart, {\it  Open Loop Stabilizability of Infinite-Dimensional
Systems}, Mathematics of Control, Signals and Systems, Vol. 11,
1998, pp. 129-160.

\bibitem {Callier_1} F. M. Callier and J. Winkin, {\it Spectral
Factorization and LQ-optimal Regulation for Multivariable
Distributed Systems}, Int. J. Control, Vol 52, No. 1, 1990, pp.
55-75.

\bibitem {Callier_2} F. M. Callier and J. Winkin, {\it LQ-optimal
Control of Infinite-dimensional Systems by Spectral
Factorization}, Automatica, Vol. 28, No. 4, 1992, pp. 757-770.

\bibitem {Staffans} O. J. Staffans, {\it Quadratic Optimal
Control through Coprime and Spectral Factorizations}, Abo Akademi
Reports on Computer Science and Mathematics, Vol. 29, 1996, pp.
131-138.

\bibitem {Weiss} M. Weiss and G. Weiss, {\it Optimal Control of
Stable Weakly Regular Linear Systems}, Math. Control Signals
Systems, Vol. 10, 1997, pp. 287-330.

\bibitem {Viola} L. Viola, E. Knill, and S. Lloyd, {\it Dynamical
Decoupling of Open Quantum System}, Phys. Rev. Lett. vol. 82,
1999, pp. 2417-2421.

\bibitem {Thorwart} M. Thorwart, P. Reimann, and P. H\"anggi, {\it
A Real-time Path Integral Method for Driven Dissipative Quantum
Systems},  Theoretische Physik I, 1999, pp. 142-144.

\bibitem {Colegrave_1} R. K. Colegrave and M. S. Abdalla, {\it A
Canonical Description of the Fabry-P\'erot Cavity }, Opt. Acta 28,
1981, pp. 495-501.

\bibitem {Colegrave_2} R. K. Colegrave and M. S. Abdalla, {\it Harmonic Oscillator with Strongly Pulsating Mass
} J. Phys. A: Math. Gen. 15, 1982, pp. 1549-1559.

\bibitem {Remaud} B. Remaud and E. S. Hernandez, {\it Damping of Wave Packet Motion
in a General Time-dependent Quadratic Field}, J. Phys. A: Math.
Gen. 13, 1980, pp. 2013-2018.

\bibitem {Kunita_1} Hiroshi Kunita,  {\it Supports of Diffusion
Process and Controllability Problems}, Proc. Intern. Symp. SDE,
Kyoto, 1976, pp. 163-185.

\bibitem {Santilli} R. M. Santilli, {\it Foundations of Theoretical Mechanics I, The Inverse Problem in Newtonian Mechanics},
Springer-Verlag, NY, 1978.

\bibitem {Tarn} T. J. Tarn, Garng M. Huang and John W. Clark, {\it Modelling of Quantum Mechanical Control Systems},
 Mathematical Modelling 1, 1980, pp. 109-121.

\bibitem {Barut} A. O. Barut and R. Raczka, {\it Theory of Group Representations and Applications}, World Scientific, 2000.

\bibitem {Kunita_2} Hiroshi Kunita,  {\it On the Controllability of
Nonlinear Systems with Applications to Polynomial Systems}, Appl.
Math. Optim., 5, 1979, pp. 89-99.

\bibitem {Isidori} Alberto Isidori, {\it Nonlinear Control
Systems}, Third Edition, Springer, 1995.

\bibitem {Clark_1} J. W. Clark, T. J. Tarn, and D. G. Lucarelli,
{\it Geometric quantum control}, Proceedings of PhysCon 2003 (St.\
Petersburg, Russia, August 20-22, 2003), in press.

\bibitem {Clark_2} J. W. Clark, D. G. Lucarelli, and T. J. Tarn,
{\it Control of Quantum Systems},  Advances in Quantum Many-Body
Theory, Vol. 6, 2002, edited by R. F. Bishop, T. Brandes, K. A.
Gernoth, N. R. Walet, and Y. Xian, World Scientific, Singapore,
pp. 411-424.

\bibitem{Chow} W. L. Chow, {\it \"Uber systeme von linearen
partiellen defferentialgleichungen erster ordnung}, Math. Ann.,
Vol. 117, 1939, pp. 98-105.

\bibitem {Hochschild} G. Hochschild, {\it The Structure of Lie Groups }, Holden-Day, San Francisco, 1965.

\bibitem {Ichihara_1} K. Ichihara and H. Kunita, {\it A Classification of
the Second Order Degenerate Elliptic Operators and its
Probabilistic Characterization}, Z.  Wahrscheinlichkeitstheorie
und Verw. Gebiete, 30, 1974, pp.\  235-254.

\bibitem {Ichihara_2} K. Ichihara and H. Kunita, {\it Supplements and
Corrections to the Above Paper}, Z. Wahrscheinlichkeitstheorie und
Verw. Gebiete, 39, 1977, pp. 81-84.

\bibitem {Lan_2} C. Lan, {\it Controllability of Time-dependent
Quantum Control Systems}, Doctoral Dissertation, Washington
University, 2003.

\bibitem{Walls} D. F. Walls and G. J. Milburn, {\it Quantum Optics
}, Springer-Verlag, Berlin, 1995.

\bibitem {Brif} C. Brif, A. Vourdas, and A. Mann, {\it Analytic
Representations Based on SU(1,1) Coherent States and Their Applications},
J. Phys. A, Vol. 29, 1996, pp.~5873-5885.

\bibitem {Gerry_1} C. C. Gerry, {\it Application of SU(1,1) Coherent
States to the Interaction of Squeezed Light in an Anharmonic
Oscillator}, Phys. Rev. A, Vol. 35, 1987, pp. 2146-2149.

\bibitem {Gerry_2} C. C. Gerry, {\it Correlated Two-mode SU(1,1) Coherent States: Nonclassical Properties}, J. Opt. Soc. Am. B, Vol.
8, 1991, pp. 685-690.

\bibitem {Zhang} L. Zhang, G. Yang, and D. Cao, {\it Generalized
Phase States and Dynamics of Generalized Coherent States}, Phys.
Lett. A, Vol. 308, 2003, pp. 235-242.

\bibitem {Boyer} C. P. Boyer, E. G. Kalnins, and W. Miller, {\it
Lie Theory and Separation of Variables 6. The Equation
$iU_t+\Delta _2 U=0$}, J. Math. Phys., Vol. 16, No. 3, 1975, pp.
499-511.

\bibitem {ZhangShengli} S. Zhang and F. Li, {\it Unitary
Transformation Approach to the Exact Solutions of Time-dependent
Quantum Systems with SU(1,1) Dynamical Group}, J. Phys. A: Math.
Gen., Vol. 29, 1996, pp. 6143-6149.

\bibitem {Leblond} J-M. L\'evy-Leblond, {\it Position-dependent
Effective Mass and Galilean Invariance}, Phys. Rev. A, Vol. 52,
No. 3, 1995, pp. 1845-1849.

\bibitem {Chetouani} L. Chetouani, L. Dekar, and T. F. Hammann,
{\it Green's Functions via Path Integrals for Systems with
Position-dependent Masses}, Phys. Rev. A, Vol. 52, No. 1, 1995,
pp. 82-91.



\end{thebibliography}
\end{document}